\documentclass[12pt]{article}
\usepackage{graphicx,lscape}

\def\lsim{\mathrel{\rlap{\lower3pt\hbox{\hskip0pt$\sim$}}
    \raise1pt\hbox{$<$}}}         
\def\gsim{\mathrel{\rlap{\lower4pt\hbox{\hskip1pt$\sim$}}
    \raise1pt\hbox{$>$}}}         

\begin{document}

\renewcommand{\theequation}{\thesection.\arabic{equation}}
\newcommand{\beq}{\begin{equation}}
\newcommand{\eeq}{\end{equation}}
\def\beqn{\begin{eqnarray}}
\def\eeqn{\end{eqnarray}}
\newcommand{\qt}{\tilde q}
\newcommand{\Tr}{{\rm Tr}\,}

\newcommand{\E}{{\cal E}}

\newcommand{\qtu}{\tilde q_{1}}
\newcommand{\qtd}{\tilde q_{2}}

\newcommand{\ax}{\vert\xi\vert}
\newcommand{\xva}{\vert\vec{\xi}\vert}
\newcommand{\rt}{\tilde r}
\newcommand{\ar}{i\theta}
\newcommand{\as}{\alpha^{2}}
\newcommand{\gs}{g^{2}}
\newcommand{\fl}{\phi_{l}}
\newcommand{\fh}{\phi_{h}}
\newcommand{\pu}{\psi_{1}}
\newcommand{\pd}{\psi_{2}}

\begin{titlepage}
\renewcommand{\thefootnote}{\fnsymbol{footnote}}

\begin{flushright}
TPI-MINN 02/12\\
UMN-TH-2052/02\\
PNPI-TH-2474/02\\
ITEP-TH-23/02\\

\end{flushright}

\vfil

\begin{center}
\baselineskip20pt
{\bf \Large Metastable Strings in  Abelian Higgs Models Embedded in
Non-Abelian  Theories: Calculating the Decay Rate}
\end{center}
\vfil

\begin{center}
{\large   M. Shifman$^a$ and   Alexei Yung$^{a,b,c}$}

\vspace{0.3cm}

$^a${\it Theoretical Physics Institute, University of Minnesota,
Minneapolis, MN 55455, USA}\\
$^b${\it Petersburg Nuclear Physics Institute, Gatchina, St. Petersburg
188300, Russia}\\
$^c${\it Institute of   Theoretical and Experimental Physics,
Moscow  117250, Russia}

\vfil

{\large\bf Abstract} 
\vspace*{.25cm}

\end{center}

We study the fate of  U(1) strings embedded
in a non-Abelian gauge theory with the hierarchical pattern of the symmetry
breaking: ${G} \stackrel{V}{\longrightarrow} {\rm U(1)}
\stackrel{v}{\longrightarrow} {\rm nothing}$,  $V\gg v$.
While in the low-energy limit the Abrikosov-Nielsen-Olesen
string (flux tube) is perfectly stable, 
being considered in the full theory it is metastable.
We consider the simplest example: the magnetic flux tubes in the
 SU(2)  gauge theory with adjoint and fundamental scalars.
First, the adjoint scalar develops a vacuum expectation value $V$
 breaking  SU(2) down to U(1). Then,
 at a  much lower scale,  the fundamental
scalar  (quark) develops  a vacuum expectation value $v$
creating the Abrikosov-Nielsen-Olesen string. 
(We also consider an alternative scenario in which the second breaking,
${\rm U(1)}
\stackrel{v}{\longrightarrow} {\rm nothing}$, is due to an adjoint field.)
We suggest an illustrative {\em ansatz} describing  an ``unwinding"
in SU(2) of the winding inherent to the Abrikosov-Nielsen-Olesen strings
in U(1). This {\em ansatz} determines an 
 effective 2D theory for the unstable mode 
on the  string world-sheet. We calculate the decay rate 
(per unit length of the string) in this {\em ansatz} and then derive a general formula.
The decay rate is exponentially 
suppressed. The suppressing exponent is
proportional to 
the ratio  of the monopole mass  squared to the 
string tension, which is quite natural in view of the
 string breaking  
through the
monopole-antimonopole pair production. We compare our result 
with the one given by  Schwinger's formula dualized
for describing the  monopole-antimonopole pair production in the 
 magnetic field.

\vfil

\end{titlepage}

\newpage

\section{Introduction}
\label{one}

Formation of the chromoelectric flux tubes in non-Abelian gauge 
theories is being discussed as a mechanism of  color confinement,
at a qualitative level,  since the early days of quantum chromodynamics (QCD).
No quantitative first-principle  description of the phenomenon
was ever constructed in QCD --- strong coupling regime inherent to this theory
precluded all efforts in this direction. 

The revival of   interest to  non-Abelian gauge theories
occurred in the mid-1990's. The idea of using supersymmetry as a tool
allowing one to deal, to a certain extent,
 with strong coupling regime was revived.
This development culminated in the work  of Seiberg and Witten
\cite{SW1,SW2} who considered ${\cal N}=2$ Yang-Mills theory
slightly perturbed by a (small) mass term  
of the adjoint matter field. This perturbation breaks
${\cal N}=2$ down to ${\cal N}=1$.
Supersymmetry proved to be sufficiently powerful
to allow Seiberg and Witten to constructively demonstrate 
the existence of the dual Meissner effect ---
the monopole condensation accompanied by
the formation of the chromoelectric flux tubes.
Technically, the Seiberg-Witten theory has two distinct scales 
(this was crucial for their construction):
the scale of strong interaction $\Lambda$, and a much smaller scale
regulated by a small adjoint mass term. Below $\Lambda$
the original  SU(2) Yang-Mills theory
reduces to an Abelian (dual) quantum electrodynamics (QED).
Correspondingly, the flux tubes (strings) of the 
Seiberg-Witten theory are the conventional U(1)
Abrikosov-Nielsen-Olesen (ANO) strings \cite{A,NO,B}. 
In the  SU($N$) case, the low-energy effective theory is that
of  U(1)$^{N-1}$; the flux tubes one deals with
present a straightforward generalization of the ANO strings.

In the Yang-Mills theory without quarks, 
the underlying gauge group has a non-trivial $\pi_1$ homotopy.
On topological   grounds one then expects $Z_N$ vortices to
appear \cite{deV,VS,HV,SS,KB,KS}. 
However, in fact, in the Seiberg-Witten construction,
in which at low energies  the gauge group is broken down to U(1)$^{N-1}$
by the  adjoint matter condensation, it is the 
  U(1)$^{N-1}$ strings that appear 
 \cite{DS} near the monopole/dyon vacua.
 They form $N-1$
infinite towers of the ANO flux tubes for each of U(1) 
factors and give rise to confinement
of quarks. It is clear that only some of these
strings (those which correspond to $Z_N$ strings of the  microscopic
non-Abelian theory) are stable, all others must be 
unstable (metastable) \cite{S} beyond the extreme low-energy limit.

A slightly different scenario takes place
in  the Seiberg-Witten theory with fundamental matter (quarks).
  In this case  the underlying gauge group
 SU($N$)  has a trivial homotopy group, $\pi_1({\rm SU}(N))=0$, and 
does not admit flux tubes.
However,  near the charge vacua this theory has an Abelian
low-energy description too,  which ensures the presence
of the ANO flux tubes in the extreme low-energy limit.
These ANO strings give rise to   confinement
 of monopoles \cite{HSZ,Y99,Y01,MY}.
It is perfectly clear that {\em all}  these  ANO strings must
be  metastable in the microscopic non-Abelian theory.

Thus, we see that the following general question presents a
 considerable interest.
Assume that one considers an underlying non-Abelian
gauge theory with a trivial (or ``almost trivial") homotopy group $\pi_1$.
(For definiteness, one may choose SU(2).)
Assume that this theory experiences a two-stage spontaneous 
symmetry breaking:
first, at a high scale $V$, the non-Abelian group
is broken down to an Abelian subgroup
(let us say, U(1), for definiteness),
and then this U(1),  in turn, is spontaneously broken
at a much lower scale $v$. 
The underlying non-Abelian theory will be referred to as ``microscopic,"
while the low-energy Abelian  theory will be called ``macroscopic."
In the low-energy limit one can forget about the microscopic
theory and consider QED with a charged matter field which
develops a   vacuum expectation value (VEV). Correspondingly, stable 
ANO strings do exist in the macroscopic
theory. In the microscopic
theory they are metastable, rather than stable, however, because of the triviality
of $\pi_1$ of the original gauge group.
A long ANO string can (and will) break due to the monopole-antimonopole pair
creation. The task is to find the probability
(per unit length per unit time) of the string breaking.

This question can be quantitatively addressed at weak coupling.
In this paper we consider a simple  
(non-supersymmetric)   model
which closely follows the pattern of the (supersymmetric)
Seiberg-Witten theory. We start from an SU(2) gauge model
with scalar fields --- one in the adjoint and another 
in the fundamental representation --- with
a  certain interaction  between them. The field in the 
fundamental representation
will be referred to as the ``quark field."
The interaction of scalars is  arranged in such a way that 
the adjoint scalar develops a large  vacuum expectation value,
\beq
V\gg  \Lambda\,,
\label{1one}
\eeq
where $\Lambda$ is the dynamical scale of the SU(2) theory.
This VEV of the adjoint field breaks the  SU(2)  gauge group down
to U(1) and ensures that the theory at hand is weakly coupled. 
At this stage the 't Hooft-Polyakov monopoles emerge \cite{thooft,polyakov}.
Their mass is very heavy, $M_M\sim V/g$.
Below scale
$V$ one is left with QED. The (charged) quark field 
 develops a small VEV $v$,
\beq
v\ll V\,.
\label{2two}
\eeq
Then the standard ANO flux tubes emerge. 
We  study their decay in the quasiclassical approximation.
To this end we consider  dynamics
of an unstable mode associated with the possibility of ``unwinding"
the ANO string
winding   on the SU(2) group manifold. This is an under-barrier process,
with the corresponding action being very large in the limit
$v\ll V$. The physical interpretation of this tunneling process
is the monopole-antimonopole pair creation accompanied
by annihilation of a segment of the string.

We present an analytic {\em ansatz} which explicitly ``unwinds" the string.
The string decay rate is found in the framework of  this  {\em ansatz}.
Our task --- constructing fully analytic  {\em ansatz} --- is admittedly illustrative.
Therefore, we limit ourselves to a minimal number of   profile functions.
The {\em ansatz} obtained in this way is suitable for a qualitative understanding
of the phenomenon.
It is too restrictive to describe the production of the monopoles {\em per se};
rather, it describes the production of a pair of highly excited states with
the monopole (antimonopole) quantum numbers. We argue, however,
that an effective description of the string decay suggested by this {\em ansatz}
is more general and is applicable for realistic {\em ans\"{a}tze} provided that
(i) the condition (\ref{2two}) is met, and (ii)
 the string tension and the monopole mass are treated  as given
parameters. In terms of these parameters the
 general result for the metastable
string decay rate (the probability per unit time per
unit length of the string) is
\beq
\Gamma_{\rm breaking}  \sim v^2\, \exp\left\{ -\, \frac{\pi \,\,  M_M^2}{T_{\rm
ANO}}\right\}\,,
\label{ura}
\eeq
where $M_M$ is the monopole  mass and  $T_{\rm ANO}$ is the string tension.
It is worth reminding that $M_M^2\sim V^2/g^2$ while $T_{\rm ANO}\sim v^2$.

 As was expected, $\Gamma_{\rm breaking}$  turns out to be exponentially suppressed. This is natural in light
of the tunneling interpretation  of the string breaking process
---  that the string is broken into pieces by the 
monopole-anti-monopole pair production.  In fact, our
result is quite similar to Schwinger's
formula \cite{Schw} dualized to  the magnetic charge production in the
constant magnetic field \cite{AfMa}. The only difference is the replacement
of $Bg^{-2}$ in the dualized Schwinger formula by  $T_{\rm ANO}$.
This is not surprising: the macroscopic theories 
of  these two phenomena are similar.

The result presented in Eq.~(\ref{ura}) is not new. It was obtained long ago
in Refs.~\cite{Vil,PV}. In these papers the decay  of metastable
ANO strings arising in theories with a hierarchical pattern of symmetry breaking
 was treated in the framework of an effective
theory,   by calculating  the action of the bubble
formed by the monopole world line on the string world sheet.

In our present paper the logic is different. We ``forget'' for a while about 
monopoles {\em per se}, and use an ``unwinding'' {\em ansatz} to derive a 
quasi-modulus  theory for the unstable string mode on the 
string world sheet. We then  consider the bubble creation in this quasi-modulus
 theory and rederive Eq.~(\ref{ura}). After this is done, we
interpret (\ref{ura}) as the probability of the monopole-antimonopole pair
production. From this standpoint   our results can be viewed as
a demonstration of the assertion that the ``unwinding'' of a metastable string goes
via the production of the monopole-antimonopole pairs.
We thus provide a necessary background for the effective
approach formulated in the pioneering works \cite{Vil,PV}.

If the small-VEV field of the quark type (i.e. in the fundamental representation)
is replaced by a small-VEV field  in the adjoint (such that the VEV's of 
two adjoint fields present in this model are misaligned), then
one arrives at a weak coupling model of $Z_2$ strings.
The gauge symmetry of the microscopic theory is now SU(2)/ $Z_2$.
Since $\pi_1 (\mbox{SU(2)/}Z_2) =Z_2$ , the  minimal magnetic flux 
string is absolutely
topologically stable. Higher-flux strings are stable   only
in the low-energy U(1) limit, and can decay through the monopole-antimonopole
pair production.
We construct an ``unwinding" {\em ansatz} in this case too,
and argue that the decay rate is given by the same expression
(\ref{ura}). 

The paper is organized as follows.
In Sect. \ref{two} we formulate our model
and explain, in concrete terms, what needs to be done
in order to calculate the string decay probability.
In Sect.~\ref{three} we present an 
{\em ansatz} for the gauge and scalar
fields describing the under-barrier transition
(``unwinding") under consideration. The decaying
string {\em ansatz} is parametrized by three profile functions.
We identify an unstable mode $\theta$ in which tunneling occurs.
For the extreme type II and type I strings one needs
to know only the asymptotic behavior of the above profile functions.
This simplifies the problem immensely.

Section \ref{four} is devoted to the extreme type II string 
which arises in the theory with the quark mass much larger then the 
photon mass
($m_q\gg m_\gamma$).
In this limit, 
the stable ANO solution was obtained analytically by Abrikosov
long ago \cite{A}. We generalize this solution to cover
 the decaying string. 
For the unstable mode $\theta$
we derive an effective two-dimensional field theory 
on the string world-sheet. The string breaking in the microscopic
theory corresponds to the false vacuum decay in the
  effective world-sheet  field theory for $\theta (t,z)$.
We then apply well-developed methods  \cite{VKO,C} for calculating the
 false vacuum decay rate through bubble formation.
In this way we find the probability of the string breaking for
the extreme type II strings.

In Sect.~\ref{five} we consider the opposite case of the extreme
type I string  ($m_q \ll m_{\gamma}$).
The analytic solution in  this case was obtained 
quite recently in Ref.~\cite{Y99}. Although the
string tension for type I is given by an expression 
significantly  different from that for type II,
 the  string  decay
rate, being expressed in terms of the
string tension,  is determined by the same formula
as in  the type II case.

In Sect. 7 we turn to the theory with two adjoint scalars
and work out the decay rate of the string with winding number
$n=2$. Note, that the string with minimal winding number $n=1$
($Z_2$-string) is stable in this model.
 Section \ref{doptwo} extends the proof
of Eq.~(\ref{ura}) to non-extreme strings.

In Sect.~\ref{six} we compare our result for the string  decay rate
 with the one given by  Schwinger's
formula \cite{Schw} which might be used for the evaluation
of the probability of 
 the monopole-antimonopole pair  production  in the external magnetic field. 
The comparison can be performed only qualitatively. 

The reason is that Schwinger's formula and Eq.~(\ref{ura})
refers to different phases of the theory.
Schwinger's formula deals with the monopole production
in the external magnetic field in the Coulomb phase, while
Eq.~(\ref{ura}) describes breaking of ANO string by
monopole-anti-monopole pair in the phase in which monopoles are confined.
Still, the comparison  exhibits a qualitative agreement with  result
(\ref{ura})
as far as the  powers of the mass scales in  the exponent  are concerned.

Finally, Sect.~\ref{seven} summarizes our results and
 conclusions and outlines problems for future investigation.
Appendix contains some details of an ``improved" 
unwinding {\em ansatz}.

\section{The model and formulation of the problem}
\label{two}
\setcounter{equation}{0}

In the bulk of the paper we will consider SU(2)  gauge theory with the action  
\beq
\label{mod}
S=\int d^4 x \left\{ \frac1{4g^2}F_{\mu\nu}^{a}F^{\mu\nu\, a} +
\frac{1}{2} (D_{\mu}\phi^a)^2 +|\nabla_{\mu}q|^2 +V(q,\phi)
 \right\}\,,
\eeq
where $\phi^a$ ($a=1,2,3$) is a real scalar  field in the  adjoint,
while $q_k$ ($k=1,\,2$) is a complex scalar  field in the  fundamental
(sometimes, we will refer to it as to the ``quark" filed). Finally, $g$ is the gauge 
coupling, and 
$V(q,\phi)$ is a scalar self-interaction potential.
Throughout the paper we will deal with  the adjoint fields
 both, in the  matrix and vector notations, say 
$$\phi\equiv \frac{\tau^a}{2}\phi^a\,.$$

The covariant derivatives $D_{\mu}$ and $\nabla_{\mu}$ act
in the adjoint and fundamental representations, respectively.
The simplest form of the potential $V(q,\phi )$ that will
serve  our purpose is  
\beq
\label{pot}
V(q,\phi)= \lambda \,  \left(|q|^2-v^2\right)^2 +\tilde{\lambda} \,
\left(\phi^a\phi^a -V^2\right)^2 +\gamma \,\left| \left(\phi-\frac{V}{2}\right) 
q \right|^2\,,
\eeq
where $v$ and $V$ are parameters of dimension of mass and
$\lambda\,, \,\, \tilde{\lambda} $ and $\gamma$ are 
dimensionless coupling constants.
In this work  we limit ourselves to  the case of weak couplings, when
all four coupling constants $g^2$, $\lambda$,
 $\tilde{\lambda}$ and $\gamma$,  are small. We also  assume that
$V\gg \Lambda$, where $\Lambda$  is the scale parameter of the SU(2) 
gauge theory. 
Then the quasiclassical treatment applies. 
Since our goal is a non-perturbative string decay, 
 we will ignore perturbative
quantum corrections altogether.

To arrange the double-scale (hierarchical) pattern of the symmetry breaking
mentioned in Sect.~\ref{one} we must ensure a hierarchy of the 
vacuum expectation values (VEV's). Namely, the breaking
SU(2)$\to$U(1) occurs at a high scale, while 
U(1)$\to$ nothing at a much lower scale,
\beq
\label{vV}
v\ll V\,.
\eeq

At the first stage the
adjoint field $\phi$  develops a VEV which can be always aligned along the 
third axis in the isospace,
\beq
\label{phivev}
\langle \phi^a
\rangle
=\delta^{a3}\, V\,. 
\eeq
This  breaks the gauge
SU(2)  group down to  U(1)  and  gives masses to the $
W^\pm$ bosons, and to one real adjoint scalar
$\phi^3$,
\beq
\label{phimass}
m_{W^\pm} =  gV\,,\qquad m_{\rm adj}\equiv m_a =
 2\sqrt{2\tilde\lambda}\,
V\, ,
\eeq
while two other  adjoint scalars ($\phi^1$ and $\phi^2$ )
are ``eaten''  up by the Higgs mechanism. Note that simultaneously the second component
of the quark field, $q_2$,    acquires a large mass,
\beq
\label{26}
M_{q_2} =\sqrt{\gamma} \, \,V\,,
\eeq
due to the last term in the potential (\ref{pot}).

Below the scales   (\ref{phimass}),  (\ref{26})  the effective low-energy
theory reduces to QED: the U(1) gauge field $A^3_{\mu}$
interacting with one complex  scalar quark $q^1$.  The action is
\beq
\label{qed}
S_{\rm QED}=\int d^4 x 
\left\{  \frac1{4g^2}\, F_{\mu\nu}^3 F^{\mu\nu\,3} 
 +\left| \tilde{\nabla}_{\mu}q_1 \right|^2 +
  \lambda \left( | q_1 |^2 -v^2 \right)^2 
\right\}\, ,
\eeq
where $$\tilde{\nabla}_{\mu}=\partial_{\mu} - \frac{i}{2} A^3_{\mu}\,.$$

Furthermore, at this second stage the charged field $q_1 $ develops a VEV, and 
the U(1) theory finds itself in 
 the Higgs phase,
\beq
\label{qvev}
\langle q_1\rangle =v\, ,\qquad (\mbox{while}\,\,\, \langle q_2\rangle =0)\,.
\eeq
At this stage the   gauge group is  completely broken. 
The breaking of U(1) gives a mass to the
photon field $A^3_{\mu}$, namely,
\beq
\label{mph}
m_{\gamma}=\frac{1}{\sqrt{2}}\,  g v\, ,
\eeq
while the mass of the light component of the quark field $q_1$
is
\beq
\label{mq}
m_{q_1}=2 \sqrt{\lambda}\,  v\,.
\eeq
In what follows we will essentially forget
about the heavy component of the quark field $q_2$, it will be irrelevant for our
consideration \footnote{The only place where
$q_2$ surfaces again, implicitly, is in  Eq.~(\ref{anz}) at $\theta \neq 0$. There is
no menace of confusion, however.}.  Only
$q_1$ is relevant. Correspondingly, in the bulk of the paper
 we will drop the subscript 1 in mentioning
the quark field; by definition, $m_q\equiv m_{q_1}=2 \sqrt{\lambda} v$.

The theory (\ref{qed}) is an Abelian Higgs model which
admits the standard Abrikosov-Nielsen-Olesen (ANO) strings \cite{A,NO}. Let us 
briefly review
 their basic features.
For generic values of $\lambda$ in Eq. (\ref{qed}) the quark mass $m_{q_1}$ (the
inverse correlation
length) and the photon mass $m_\gamma$ (the
inverse penetration depth) are distinct.
Their ratio is an important parameter in the theory of superconductivity,
characterizing the type of
superconductor. Namely, for $m_{q_1} < m_\gamma$ one deals with the type I superconductor
in which two strings at large separations attract each other. On the other hand,
  for $m_{q_1} >m_\gamma$ the superconductor is of type II, 
in which two strings at large separations repel each other.
 This  behavior is related to the fact
that  the scalar field generates  attraction between two vortices,
while the electromagnetic field  generates repulsion.
The boundary separating superconductors of
the I and II types  corresponds to $m_{q_1} =m_\gamma$, i.e. to
a special value of the quartic coupling $\lambda$, namely,
\beq
\lambda    = \frac{g^2}{8} \,.
\label{211}
\eeq
In this case the vortices do not interact.

It is well known that the point (\ref{211})
represents, in fact, the Bogomolny-Prasad-Sommerfield (BPS) limit.
At $m_{q_1}=m_\gamma$ the ANO string 
satisfies  first order differential equations and saturate
the Bogomolny bound~\cite{B}. In supersymmetric theories
the Bogomolny bound for the BPS strings coincides with the value
of the central charge of the SUSY algebra \cite{HS,DDT,GS}.
In particular, the BPS strings arise in the Seiberg-Witten theory
near the monopole/charge vacua at small values of the adjoint
mass perturbation \cite{HSZ,FG,VY}.

As was mentioned in Sect.~\ref{one},
we will also consider,  in brief, a model with two adjoint matter fields,
\beq
\label{modprime}
S=\int d^4 x \left\{ \frac1{4g^2}F_{\mu\nu}^{a}F^{\mu\nu\, a} +
\frac{1}{2} (D_{\mu}\phi^a)^2 + \frac{1}{2} (D_{\mu}\chi^a)^2 +U(\chi ,\phi)
 \right\}\,,
\eeq
where $\phi^a$ and $\chi^a$ ($a=1,2,3$) are real scalar  fields, and the
potential $U(\chi ,\phi)$ can be chosen, for instance, as follows:
\beq
\label{potprime}
U(\chi ,\phi) = 
\alpha \left(\phi^a\phi^a -V^2\right)^2
+\frac{\beta}{4}\,  \left(\chi^a\chi^a -2 v^2\right)^2
+\gamma\left(\phi^a\chi^a \right)^2
\eeq
The condensation of the $\phi$ field, Eq.~(\ref{phivev}),
breaks the gauge symmetry SU(2)/$Z_2\to$ U(1), and 
makes $W$ bosons, $\phi^3$ and $\chi^3$ heavy. What 
remains in the low-energy limit?
The low-energy is very similar to (\ref{qed}), namely,
\beq
\label{qedprime}
S_{\rm QED'}=\int d^4 x 
\left\{  \frac1{4g^2}\, F_{\mu\nu}^3 F^{\mu\nu\,3} 
 +\left( D_{\mu}\chi^+ \right) \left( D_{\mu}\chi^- \right) +
  \beta \left(  \chi^+\chi^- -v^2 \right)^2 
\right\}\, ,
\eeq
where 
\beqn
\chi^\pm &\equiv& \frac{1}{\sqrt 2}\left(\chi^1\pm i\chi^2
\right)\,,\nonumber\\[2mm]
D_{\mu}\chi^\pm &=& \left( \partial_{\mu} \pm i A^3_{\mu}\right)\chi^\pm\, .
\label{frione}
\eeqn
At the second stage the remaining U(1) is completely broken by the condensate of the
$\chi$ field,
\beq
|\langle \chi\rangle | = v\,.
\eeq
This gives mass to the ``photon" field $A^3$, and the ANO strings enter the game.

\section{Abrikosov-Nielsen-Olesen String}
\label{dopone}
\setcounter{equation}{0}

\begin{figure}[htb]
\begin{center}
\includegraphics[width=9cm]{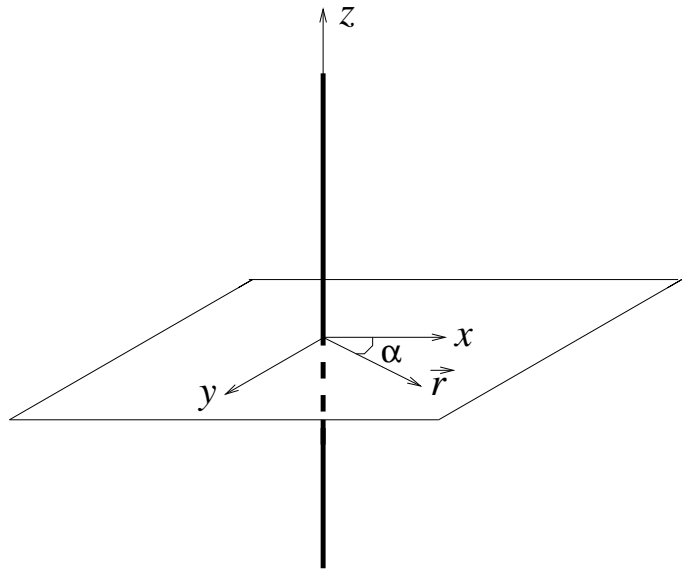}
\end{center}
\caption{
Geometry of the string. }
\label{suf1}
\end{figure}

In the model (\ref{qed})
the classical field equations for the  ANO string with the unit  winding number 
are solved in  the standard {\em ansatz},
\begin{eqnarray}
q_1(x) &=& q (r)\, {\rm e}^{- i\,\alpha}\, ,\nonumber\\[2mm]
A^{3}_0 &\equiv& 0\, ,\nonumber\\[2mm]
  A^{3}_i(x) &=& 2\epsilon_{ij}\,\frac{x_j}{r^2}\left[1-f(r)\right]\,.
\label{anoprof}
\end{eqnarray}
Here $$r=\sqrt{\sum_{i=1,2} x_j^2}$$ is the distance from
the vortex center  while $\alpha$
 is the polar angle in the transverse to the  vortex axis  
$(1,2)$-plane (the subscripts $i,j=1,2$ denote coordinates in this plane, $x$ and $y$, 
see Fig.
\ref{suf1}).
Moreover, $q(r)$ and $f(r)$ are profile functions.
Note, that $\partial_{i}\alpha=-\epsilon_{ij}\, {x_j}/{r^2}$.

 The profile functions $q$ and $f$ in Eq.~(\ref{anoprof}) are real and
satisfy the second order differential equations
\begin{eqnarray}
q''  &+& \frac1{r}q'-\frac1{r^2}f^2 q-m_{q}^2\,\, \frac{q(q^2-v^2)}{2v^2}=0 
\, ,\nonumber\\[2mm]
f'' &-& \frac1{r}f'-\frac{m_{\gamma}^2}{v^2}\,
q^2 f=0 
\label{anoeq}
\end{eqnarray}
for generic values of $\lambda$ (the prime stands here for the
 derivative with respect to $r$),
plus the boundary conditions

\newpage 

\begin{eqnarray}
&& q (0)=0\ ,
\qquad ~f(0)=1 
\, ,\nonumber\\[2mm]
&&  q (\infty)=v\ , \quad
f(\infty)=0\,,
\label{bc}
\end{eqnarray}
which ensure  that the scalar field reaches its VEV ($q_1=v$) at
infinity and the vortex at hand carries one unit of magnetic flux.

 The expression
for the string tension (energy per unit length) for the ANO string 
in terms of the profile functions (\ref{anoprof}) has the form
\beq
T_{\rm ANO}=2\pi \int rdr
\,
\left\{
\frac{2}{g^2}\, \frac{f'^2}{r^2} +q'^2
+\frac{f^2}{r^2}q^2 +\lambda (q^2-v^2)^2
\right\}\,.
\label{anoten}
\eeq

For generic values of the ratio $m_q/m_{\gamma}$  only a numerical
solution of Eqs.~(\ref{anoeq}) is possible.
 However, in the extreme type II
case ($m_q\gg m_{\gamma}$) and extreme type I case
($m_q\ll m_{\gamma}$) analytical solutions
can be readily  found   \cite{A,Y99}. We will review 
these solutions in 
Sects. \ref{four} and \ref{five}, respectively. As was explained in Sect. \ref{one},
in the full SU(2) theory the ANO string can decay. 
We will use the solutions discussed in
Sects. \ref{four} and \ref{five} in order to analytically
calculate the  decay rate of the ANO string. 

The magnetic field flux for the string (\ref{anoprof}) is
\beq
\frac{1}{2}\, \int B^3 dx\,dy \equiv \frac{1}{2}\, \oint A^3_i dx_i = 2\pi\,.
\label{fritwo}
\eeq

In the model with two adjoints, Eq.~(\ref{qedprime}),
the ANO string solution has the form
\begin{eqnarray}
\chi^-  &=&  h_{1/2} (r)\, {\rm e}^{- i\,\alpha}\,
 ,\qquad h(0)_{1/2}=0\,,\qquad h_{1/2}(\infty )
= v\,,
\nonumber\\[2mm] A^{3}_0 &\equiv& 0\, ,\nonumber\\[2mm]
  A^{3}_i(x) &=& \epsilon_{ij}\,\frac{x_j}{r^2}\left[1-f(r)\right]\,,
\qquad f(0) = 1\,,\qquad f(\infty ) = 0\,.
\label{anoprofprime}
\end{eqnarray}
In this case
\beq
 \int B^3 dx\,dy \equiv   \oint A^3_i dx_i = 2\pi\,.
\label{frithree}
\eeq
The magnetic flux is twice smaller.
That's the reason why the elementary string in the model
(\ref{qedprime})
cannot be broken by the monopole pair production.
However, the double-winding string
\beqn
\chi^-  &=&  h (r)\, {\rm e}^{- i\,2\, \alpha}\, ,
\nonumber\\[2mm] A^{3}_0 &\equiv& 0\, ,\nonumber\\[2mm]
  A^{3}_i(x) &=& 2\epsilon_{ij}\,\frac{x_j}{r^2}\left[1-f(r)\right] 
\label{anoprofpr}
\eeqn
can and will be broken.

\section{Decaying strings:   a  `` primitive" unwinding \\
ansatz}
\label{three}
\setcounter{equation}{0}

\begin{figure}[htb]
\begin{center}
\includegraphics[width=8cm]{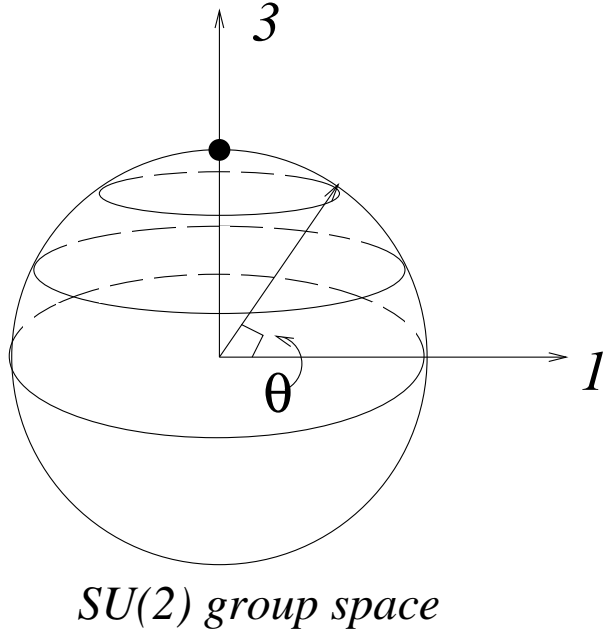}
\end{center}
\caption{Unwinding the ANO ansatz.}
\label{vstone}
\end{figure}

It is clear that the ANO strings (\ref{anoprof}) are topologically stable only
at low energies when the SU(2)  theory (\ref{mod}) reduces to a ``macroscopic" theory,
 QED, see Eq. 
(\ref{qed}). 
In the full ``microscopic'' theory (\ref{mod}) they should be
metastable because the SU(2)  gauge group does not admit flux tubes,
$\pi_1 (SU(2))=0$. To visualize the decay possibility, note that the winding in
(\ref{anoprof}) runs along the ``equator'' of the  SU(2)  group
space (which is $S_3$) and, therefore, can be shrunk  to zero by contracting the 
loop towards the south or north poles (Fig. \ref{vstone}).

It is not difficult to devise an {\em ansatz}
encoding the possibility of unwinding the field configuration
(\ref{anoprof}) through the loop 
 shrinkage  in the SU(2) group space.
The {\em ansatz} which does the job and  eventually
will allow us to calculate 
the  ANO string  decay rate is parametrized by an angle parameter $\theta$,
\begin{eqnarray}
q_{k}(x) &=&
 U\,e_{k}\, q_{\theta} (r)\, ,\qquad (e_1=1,\; e_2=0)\,,
 \nonumber\\[2mm]
A_0 &\equiv& 0\,, \qquad A_3  \equiv  0\,, \nonumber\\[2mm]
 A_j(x)   &=& i \, U\partial_{j}\, U^{-1} \, [1-f_{\theta}(r)]\, , \qquad j=1,2\,,
 \nonumber\\[3mm]
\phi &=& V\,U\, \frac{\tau_{3}}{2}\, U^{-1} +\Delta\phi \, ,
\label{anz}
\end{eqnarray}
where 
\beq
\label{matrix}
U=e^{-i\alpha\tau_3}\cos{\theta} +i\, \tau_1\,  \sin{\theta}\,.
\eeq
(Eventually, upon quantization, $\theta$
will become a slowly varying function of $z$ and $t$, a field $\theta (t,z)$.)

The gauge and quark fields in (\ref{anz}) are parametrized by
profile functions $f_{\theta}(r)$ and $q_{\theta} (r)$ 
 depending on the parameter $\theta$. They satisfy the  same boundary
conditions
\begin{eqnarray}
 &&q_{\theta} (0)=0\ ,
\qquad ~f_{\theta}(0)=1,
\nonumber\\[3mm]
&&  q_{\theta} (\infty)=v\ , \quad
f_{\theta}(\infty)=0\
\label{Aqbc}
\end{eqnarray}
as in the U(1)  case, see Eq.~(\ref{bc}). The boundary conditions at zero
are chosen  to ensure the absence of singularities
of our {\em ansatz}  at $r=0$. The magnetic flux of the string 
(\ref{anz}) is
\beq
\label{flux}
2\pi \cos^2{\theta}.
\eeq
It equals to $2\pi$ at $\theta=0$ and goes to zero
at $\theta=\frac{\pi}{2}$.

The term $\Delta\phi $ in the last line of Eq. (\ref{anz})
is needed to make sure that 
there is no singularity at $r\to 0$.
For axially symmetric string the function $\Delta\phi$
 can be chosen in the form
\beq
\Delta\phi=\varphi_{\theta}(r)\,\left[
\frac{\tau_{1}}{2}\, \sin{\alpha}-
\frac{\tau_{2}}{2}\, \cos{\alpha}\right]
\,,\label{deltaphi}
\eeq
where we assume that the component of $\Delta\phi$ along
$\tau_3$ is zero, while $\varphi_{\theta}(r)$ is an extra
profile function, which depends on $\theta$ as a parameter.
 The $a=1,2$ components of $\Delta\phi$ 
cannot be put to zero. To see this substitute Eqs. (\ref{matrix})
and (\ref{deltaphi}) into the last line in Eq. (\ref{anz}). Then one gets   
\beq
\label{phianz}
\phi=\frac{\tau_{3}}{2}\, V \, \cos{2\theta}-
\left[
\frac{\tau_{1}}{2}\, \sin{\alpha}-\frac{\tau_{2}}{2}\, \cos{\alpha}
\right]\,
\left[ V\sin{2\theta}-\varphi_{\theta}(r)\right]\, .
\eeq
From this expression it is clear that $\phi $ has no singularity at
$r=0$ provided that
\beq
\label{0phibc}
\varphi_{\theta}(0)= V\sin{2\theta}\,.
\eeq
The boundary condition for $\varphi_{\theta}(r)$ at infinity
should be chosen as follows:
\beq
\label{infphibc}
\varphi_{\theta}(\infty)=0\, .
\eeq
Both boundary conditions are consistent
with the initial condition
\beq
\left.
\varphi_{\theta} (r)\right|_{\theta =0} = 0\,,
\eeq
to be imposed.

For future reference it is convenient to present the very same {\em ansatz}
(\ref{anz}) in the singular gauge,
\beqn
q_{k}(x)  &=&  e_{k}\, q_{\theta} (r)\, ,
\nonumber\\[2mm]
A_i(x) &=& -i \,( \partial_{i}U^{-1})\, U \,  f_{\theta}(r)\, ,
\nonumber\\[2mm]
\phi &=& V\, \frac{\tau_3}{2} \, +U^{-1}\,( \Delta\phi )\,  U\, .
\label{singan}
\eeqn

Our {\em ansatz}, Eqs. (\ref{anz}) or (\ref{singan}), plus (\ref{matrix}), 
smoothly 
 interpolates between the ANO-type winding along the equator 
at $\theta= 0$, and constant matrix with no winding  at
$\theta= {\pi}/{2}$.  In other words, we start from the ANO
string at $\theta=0$ and arrive at empty vacuum at $\theta=\pi /2$.
Indeed, at  $\theta= 0$ the adjoint field $\phi$
is aligned, $\phi^a = V\delta^{3a}$, while $U=e^{-i\alpha\tau_3}$.
At $\theta={\pi}/{2}$ again $\phi^a = V\delta^{3a}$,
and $U=i\tau_1$, implying $A_j =0$. 
(Equivalently, one could have chosen to end up with 
$U=i\tau_2$ at $\theta=\frac{\pi}{2}$.) Thus, we managed to unwind 
the ANO winding through  extra dimension of the 
vacuum manifold in the
SU(2) theory which was not there in QED.

We pause here to make an additional comment  regarding our 
{\em ansatz}  (\ref{anz}). 
At large $r$, when $q_\theta\to v$ and $f_\theta\to
0$ and $\varphi_{\theta}\to 0$, our field configuration
presents a 
  gauge-transformed ``plain vacuum".  This ensures 
that at every given $\theta$ the energy functional converges
at large $r$. The convergence of the  energy functional
at small $r$ is guaranteed by the boundary conditions
$q_{\theta} (0)=0\,,
\,\,\,
f_{\theta}(0)=1$ and (\ref{0phibc}).

The tension of the string (or the field configuration in which it
evolves at $\theta \neq 0$)
is a functional of three
functions $f_{\theta}(r)$ , $q_{\theta} (r)$ 
and $\varphi_{\theta}(r)$.
It is easy to get this functional by
  substituting the {\em  ansatz} (\ref{anz})
in  the action (\ref{mod}). Restricting ourselves to (1,2)-plane we obtain
after some algebra
\begin{eqnarray}
T(\theta)
&=&\!\!\!
2\pi \int rdr\,\left\{
\frac{2}{g^2}\, \frac{f_{\theta}'^2}{r^2}\,
\cos^2{\theta}
 +q_{\theta}'^2
+\frac{f_{\theta}^2}{r^2}
\,
q_{\theta}^2 \, \cos^2{\theta}
\right.
\nonumber\\[3mm]
&+&\!\!\!
\frac{1}{2}\varphi_{\theta} '^2 +\frac1{2r^2}\left[
\varphi_{\theta}(\cos{2\theta}-
2f_{\theta}\cos^2{\theta})+Vf_{\theta}\sin{2\theta}\right]^2
\nonumber\\[3mm]
&+&\!\!\!
\left.
\lambda (q_{\theta}^2-v^2)^2+
\tilde{\lambda}\left[
\varphi_{\theta}^2
-2V\varphi_{\theta}\sin{2\theta}\right]^2 
+\frac{\gamma}{4}
\varphi_{\theta}^2q_{\theta}^2\right\}\, .
\label{ten}
\end{eqnarray}

Needless to say that at $\theta=0$ the string tension $T$
  coincides with that for the
ANO string, see (\ref{anoten}), while at $\theta= {\pi}/{2}$
it goes to zero, as was expected.

Now we have to minimize the string functional (\ref{ten})
with respect to three
profile functions
$f_{\theta}(r)$ , $q_{\theta} (r)$ 
and $\varphi_{\theta}(r)$ at fixed $\theta$. This procedure would give 
us  a solution for the profile functions. 
Finding the full solution is a rather complicated task
requiring numerical computations which go  beyond the 
scope of the present paper. However, we will 
be able to get sufficient insight in order to derive
the general formula (\ref{ura}) by purely analytical means.

Why we call the {\em ansatz} (\ref{anz}) primitive? It has only one profile function
$f_\theta (r)$ for all three gauge components. This is okay at $\theta =0$
when the $W$ boson degrees of freedom are not excited.
At $\theta \neq 0$ the universality of $f_\theta (r)$ forces one 
and the same spread  in the perpendicular
plane of the $W$ boson and photon 
components of the solution. As a result, unwinding of the ANO strings in Eq. 
(\ref{anz}) proceeds via the production of highly excited ``monopoles."  This
shortcoming can be eliminated through introduction of extra profile functions, see
Sect. \ref{doptwo} and Appendix.

\section{Extreme type II string}
\label{four}
\setcounter{equation}{0}

As we have already mentioned,
even for the stable ANO string  the  analytic solution for generic values of 
the ratio $m_{q}/m_{\gamma}$ is absent. 
On the other hand, it is not difficult to
find the solution that would be valid in the
logarithmic approximation,
$\ln m_{q}/m_{\gamma}\gg 1$.
Therefore, in this section we consider
the extreme type II superconductor for which 
we will be able to obtain an analytic
solution of the problem of  the decaying string.

In fact, we impose the following relation between the  adjoint scalar, quark and
photon masses:
\beq
\label{typeI}
m_{a}\gg m_{q}\gg m_{\gamma}.
\eeq
In addition, it is convenient to limit ourselves to the case $m_W\sim m_{a}$.
At first, we review   Abrikosov's solution for the ANO string \cite{A}, then
consider the decaying string and, finally, work out the 
effective action for the unstable mode on the string world-sheet
and calculate the string decay rate.

\subsection{Type II ANO string}
\label{arcone}

Assume that $m_{q}\gg m_{\gamma}$ in the Abelian
Higgs model (\ref{qed}).
Then the ANO string looks as follows. The quark field $q(r)$ varies from zero
to its vacuum value $v$ inside a small core of radius of order of $1/m_{q}$, whereas
the electromagnetic field is spread over a much larger domain,
of order of $1/ m_{\gamma}$. In 
the latter domain the quark field is already very close
to its VEV. The solution of the second equation in 
(\ref{anoeq}) for  the gauge profile function $f$ is $f\sim r^2$ in this 
domain (where the last term in this equation can be ignored).
Being  properly normalized, the solution has the following asymptotics:
\beq
\label{f}
f=\left\{
\begin{array}{l}
1-c\, m_{\gamma}^2 \, r^2  \quad \mbox{at}\quad  r\ll  m_{\gamma}^{-1}\,,\\[3mm]
0  \quad \mbox{at}\quad  r\gg m_{\gamma}^{-1}\,,
\end{array}
\right.
\eeq
where $c$ is a constant, $c\sim 1$. The 
approach to zero at $r\gg m_{\gamma}^{-1}$
is exponentially fast.

The leading (and the only)
 logarithmic contribution to the string tension comes from
the third term in the expression (\ref{anoten}) for the 
ANO string tension. In the domain
$ m_{q}^{-1} \ll r\ll   m_{\gamma}^{-1}$,
with the logarithmic accuracy, 
one can  substitute in Eq.~(\ref{anoten}) $q= v$ and
retain  only the leading term in the gauge profile function,
$f= 1+\cdots $. In this way one gets
\beq
T_{\rm ANO}^{\rm II}=2\pi v^2\,  \ln\, \frac{m_{q}}{m_{\gamma}}\,.
\label{aten}
\eeq
Domains other than $ m_{q}^{-1} \ll r\ll   m_{\gamma}^{-1}$
yield a non-logarithmic contribution.
The same is valid with regards to
the first, second and fourth terms in Eq.~(\ref{anoten}),
as well as deviations from  $q=v$ and $f=1$.
All these effects give 
corrections to (\ref{aten}) suppressed by powers of
 $1/ \ln (m_{q}/ m_{\gamma})$.

This result was obtained by Abrikosov in 1957 \cite{A}.
Note that the leading logarithmic contribution to
the string tension (\ref{aten}) is totally 
insensitive to details of the
profile functions
$q$ and $f$. It ``feels'' only the boundary values
$q(\infty)=v$ and $f(0)=1$.

\subsection{Decaying type II string}
\label{dt2s}

Now we use the same method of separating distinct physical
scales to describe the decaying extreme type II  string.
Our goal is calculating the barrier (potential energy versus $\theta$)
which will be later used in the calculation of the decay rate.
We will need the kinetic term for the field $\theta (t,z)$ too, but we begin
with the
potential term.

The condition (\ref{typeI}) ensures that both, the adjoint scalar and quark
fields vary in   small cores with sizes of order $1/m_{a}$ and $1/m_{q}$,
respectively. In the domain
$m_{q}^{-1} \ll r\ll   m_{\gamma}^{-1}$ the fields $\phi$ and $q$
 already reach their boundary
values $\varphi_{\theta}(\infty)=0$ and  $q_{\theta}(\infty)=v$.
Moreover, we again use Eq.~(\ref{f}) for the
 function $f_{\theta}$,
 the leading logarithmic contribution to the tension
coming from the boundary value $f_{\theta}(0)=1$. 

Substituting this data in Eq.~(\ref{ten}) we get
\beq
\label{tenII}
T^{\rm II}(\theta)= 2\pi \,  \left( \ln\, \frac{m_{q}}{m_{\gamma}}\right)
\, v^2 \, \cos^2{\theta}+
2\pi \,  \left(  \ln \, \frac{m_{a}}{m_{\gamma}}\right)
\,\frac{V^2}{2}\,
\sin^2{2\theta}\,.
\eeq

\begin{figure}[htb]
\begin{center}
\includegraphics[width=15cm]{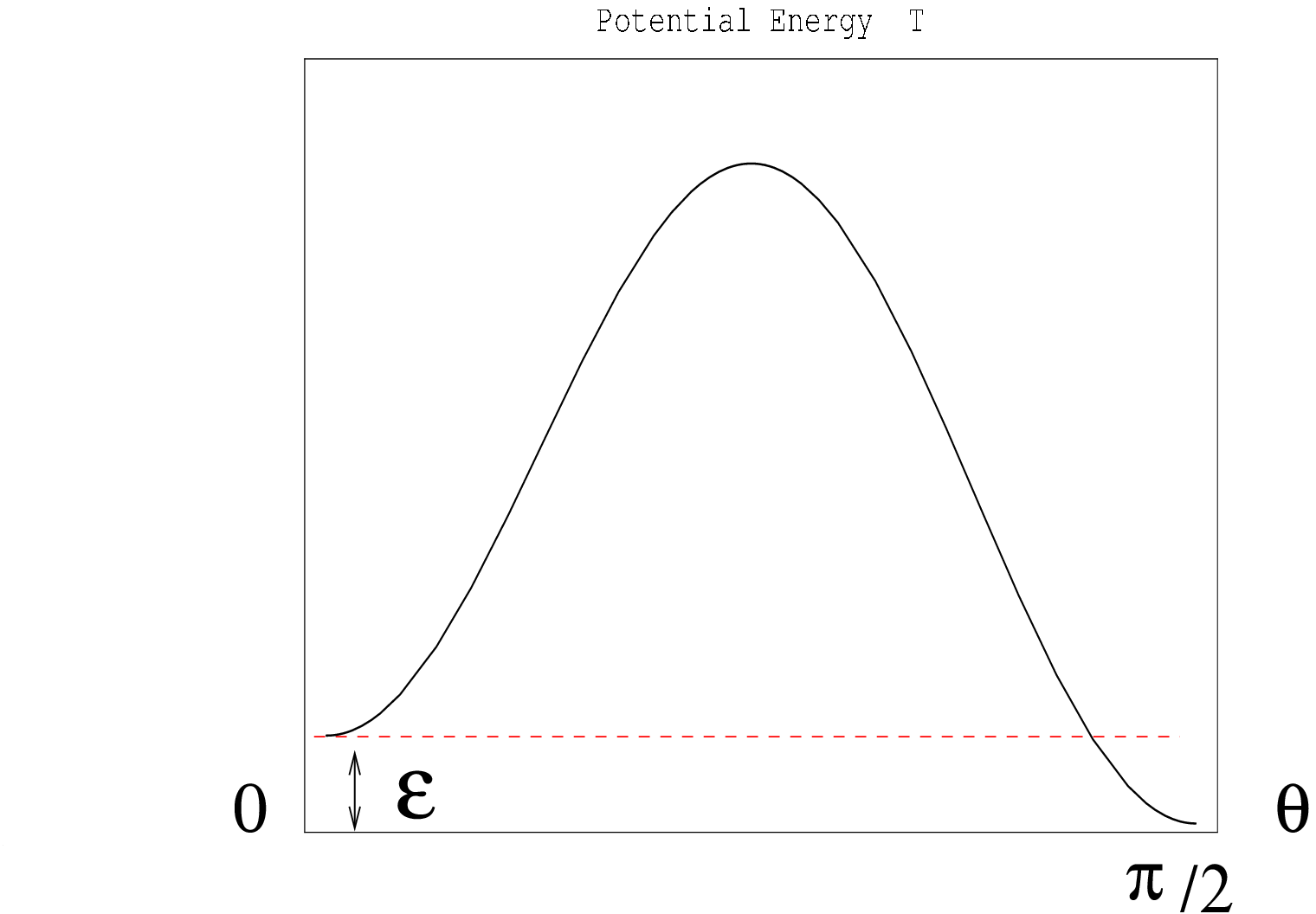}
\end{center}
\caption{
The potential energy  $T (\theta)$.}
\label{suf3}
\end{figure}

This is our result for the barrier profile for the extreme type II string.
The first term here comes from the third term in Eq.~(\ref{ten})
while the second one comes from the fifth term in Eq.~(\ref{ten}).
All other contributions to $T$ contain no large logarithms.
At $\theta=0$ the tension in (\ref{tenII}) coincides 
with that of the ANO string, see Eq.~(\ref{aten}). At non-zero $\theta$ the $V^2$
term dominates --- it produce a huge barrier, with the height of order
of $V^2\, \ln\,  ( m_{a}/m_{\gamma})$. At
$\theta= {\pi}/{2}$ the tension $T^{\rm II}(\theta)$ vanishes,
which means that  the string disappears. This is summarized in Fig. \ref{suf3}
presenting $T^{\rm II}(\theta)$ {\em versus} $\theta$.

\subsection{Effective world-sheet theory}
\label{43}

In order to calculate the decay rate of the string we have
to work out the effective theory for the unstable mode
parametrized by $\theta$ on the string world-sheet.
The collective coordinate $\theta$ becomes a field of a 
2D sigma model on the string world-sheet. The tension (\ref{tenII})
gives us the potential term in the action of this sigma model.
To complete the problem we need  to know the kinetic term.

In order to obtain the kinetic term we apply the standard strategy:
 we assume that $\theta$ adiabatically depends on the 
world-sheet coordinates $\sigma$. (Throughout the
paper we use here the static gauge
for  the string in which $\sigma_1=t$ and $\sigma_2=x_3$.)
The kinetic term for $\theta$ in the  2D sigma model  comes from those
in the action (\ref{mod}). 
To calculate the kinetic term for $\theta$
it is convenient to  rotate our field
configuration (\ref{anz}) into a ``singular'' gauge
performing the gauge transformation with the matrix $U^{-1}$,
see Eq. (\ref{singan}).
In this gauge the boundary values of the  fields at infinity
do not depend on $\theta$ and give no contribution
to the kinetic energy term, for instance, $q_{k}(x)  = 
 e_{k}\, q_{\theta} (r)$,
and
\beq
\phi = V\, \frac{\tau_3}{2} \, +U^{-1}\,( \Delta\phi )\,  U\, ,
\label{singanz}
\eeq 
(remember, $ \Delta\phi$ falls off at infinity).
The gauge field falls off rapidly at large distances,
\beqn
A_i(x) &=  -i \,( \partial_{i}U^{-1})\, U \,  f_{\theta}(r)\,.
\label{singanzz}
\eeqn

We pause here to discuss a subtle point in the calculation of
the kinetic term. With time dependence switched off, for the
static string, the non-vanishing components of the gauge potential are
$A_j$, ($j=1,2$). The components $A_n$ with $n=0,3$ vanish, see Eq.
(\ref{anz}).  However, as soon as we allow $\theta$ to depend on the 
world-sheet coordinates, the components  $A_n$ with $n=0,3$ must become
non-zero. Indeed, let us first assume that $A_n = 0$
(we will immediately see that this is a wrong assumption).
Consider  the contribution of the kinetic term of the gauge
field $F_{\mu\nu}^2$. It is clear that the only $F_{nj}^2$ piece
contributes to the kinetic term of $\theta$,
\beq
F_{n\, j}=\partial_{n}A_{j}- \partial_{j}A_{n} -i\, [A_n\, , A_j]\,,
\quad j=1,2\,,\quad n=0,3\,.
\label{Fni}
\eeq
If the components $A_n$ vanished, then one would obtain
\beq
\label{parA}
F_{n\, j}= \partial_{n}\, A_{j}= (\partial_{n}\theta )\, (\partial_{\theta}A_{j})\,,
\eeq
which, being combined with Eq. (\ref{singanz}) for the gauge potential $A_j$,
yields
\beq
\partial_{n}A_{j}= 2\, (\partial_{n}\theta) \, 
\varepsilon_{j\ell}\,
\frac{x_\ell}{r^2}\, 
\left[ \sin{2\theta}\,\, \frac{\tau_3}{2} +
\cos{2\theta}\left(\frac{\tau_1}{2}\sin{\alpha}
+\frac{\tau_2}{2}\cos{\alpha}\right)\right]f_{\theta}(r)\, .
\label{diA}
\eeq
With this formula the expression for
$F_{nj}^2$ is not even   gauge invariant with respect 
to the  gauge transformations
depending on $\theta$,  see Eq.~(\ref{parA}). 

The fact that we made a mistake by assuming
$A_n = 0$
 manifests itself in
the singularity in Eq.~(\ref{diA}) at $r=0$ (note that $f_{\theta}(0)=1$, 
see Eq.~(\ref{Aqbc})). 

Thus,  in the calculation of the kinetic term of
$\theta$ one cannot avoid switching on the
components $A_n$ with $n=0,3$ which, naturally,
must be proportional to $\partial_{n}\theta$.
The following
{\em ansatz} for $A_n$ goes through:
\beq
\label{An}
A_n=-2\,(\partial_{n}\theta)\left(\frac{\tau_1}{2}\, \cos{\alpha}
-\frac{\tau_2}{2}\, \sin{\alpha}\right)a_{\theta}(r)\, ,
\eeq
where $a_{\theta}(r)$ is a new 
 profile function and the angle $\alpha$ is defined in Fig.~\ref{suf1}. Generally
speaking,
$A_n$ is parametrized by three distinct profile function accounting for
three generators of  SU(2).  However, as it turns out,
the single structure presented in Eq.~(\ref{An}),
leads to a fully self-consistent and complete {\em ansatz},
with the $r\to 0$ singularity in $F_{nj}$  cancelled.
We  do not need two other  structures.

Substituting Eq.~(\ref{An}) in Eq.~(\ref{Fni}) and
ignoring terms that are {\em a priori}  non-singular at $r=0$,  we get
\beqn
F_{n\, j}=2\,(\partial_{n}\theta)\, \varepsilon_{j\ell }\, \frac{x_\ell}{r^2}\, 
\left[ \sin{2\theta}\,\frac{\tau_3}{2} +
\cos{2\theta}\left(\frac{\tau_1}{2}\sin{\alpha}
+\frac{\tau_2}{2}\cos{\alpha}\right)\right]\left( 1-a_{\theta}(r)
\right)\, ,
\label{Fnia}
\eeqn
where we  assume that the gauge profile function $f_{\theta}= 1+\cdots $
at $r\ll 1/m_{\gamma}$. The reason why we can keep only the  
most singular
terms is as follows.
Let us remind that in Sect.~\ref{dt2s} we found the potential term for the
$\theta$ field in the logarithmic approximation (i.e. the approximation 
in which
only those terms are kept which contain large logarithms of the
mass ratio).  Our task in this section is to find
the kinetic term for the
$\theta$ field in the very same logarithmic approximation.

In order to cancel the actual divergence of
$\int d^2 x\, F_{n\, j}^2$  at $r\to 0$
we must impose the following 
boundary condition:
\beq
\label{abc}
a_{\theta}(0)=1\,,
\eeq
as well as 
\beq
\label{abcinf}
a_{\theta}(\infty )=0\, .
\eeq
In fact, for our purposes --- determination of the
kinetic term with the logarithmic accuracy ---
it is sufficient to use the step function
model for
$a_{\theta}(r)$ similar to that in Eq.~(\ref{f}),
\beq
\label{atheta}
a_{\theta}(r)
=\left\{
\begin{array}{l}
1  \quad \mbox{at}\quad  r\ll  R\,,\\[3mm]
0  \quad \mbox{at}\quad  r\gg R\,,
\end{array}
\right.
\eeq
where we introduced a new parameter $R$, to be determined below,
(it is assumed that $R\ll 1/m_{\gamma}$).

With the profile function $a_{\theta}(r)$  presented in Eq.~(\ref{atheta}), the
contribution of the gauge term  to the kinetic energy of the field $\theta$ takes
the form
\beq
\label{king}
\int d^2 x\,  \frac1{4g^2}\, F_{\mu\nu}^a   F^{\mu\nu\,\, a} \rightarrow 
\frac{4\, \pi}{\gs} \,   (\partial_{n}\theta)^2
\,  \ln{ \frac{1}{Rm_{\gamma}}} \, .
\eeq

This is not the end of the story, however, since, in addition, we have
to take into account  the kinetic energy coming from the 
scalar kinetic terms in Eq.~(\ref{mod}) (the second and the third terms). 
It is obvious that the third term, associated with the quark field,
is proportional to $v^2$ and can be neglected as 
  compared to the second term  --- the contribution of the
 adjoint scalar which is proportional to $V^2$.
As a result, 
using Eq.~(\ref{An}) to calculate
$D_{n}\phi$, we get for the total kinetic energy
\beq
\label{kin}
4\pi \, (\partial_{n}\, \theta)^2\left[ \frac1{g^2}\ln{\frac{1}{R\, m_{\gamma}}}+
{\rm const}\cdot (V^2\,  R^2) \right]\, .
\eeq
The parameter $R$ can now be determined from the requirement
that the coefficient in front of $(\partial_{n}\theta)^2$ be minimal.
Minimizing this with respect to $R$ we find the condition
$$
\frac{1}{g^2\, R} \sim V^2\, R\,.
$$
In other words, $R$ turns out to be small, of the order of the inverse mass
of the 
 W boson,
\beq
\label{R}
R\sim \frac{1}{m_{W}}\, ,
\eeq
in full accord with what our physical intuition demands.
With this value of the parameter $R$, the first logarithmic term in Eq.~(\ref{kin})
dominates  over the second one, which can be thus ignored with the logarithmic
accuracy. As a result, combining together the kinetic term (\ref{kin}) with the
potential term (\ref{tenII}),  we  arrive at the following action of the 2D sigma
model:
\beqn
S_{\rm str\,\, II} &=& 2\pi \, \int d^2 \sigma\,
\left\{\frac{2}{\gs}\left( \ln {\frac{m_{W}}{m_{\gamma}}}\right)
\,(\partial_{n}\theta)^2 \right.
\nonumber\\[3mm]
&-&\left. \! \! \! \! 
\left( \ln{\frac{m_{q}}{m_{\gamma}}}\right) \, v^2 \cos^2{\theta}
-
\left( 
\ln{\frac{m_{a}}{m_{\gamma}}}\right)
\,\frac{V^2}2\sin^2{2\theta}\right\}\, .
\label{sigmod}
\eeqn

\vspace{2mm}

This is our final result for the effective theory
of the unstable $\theta$-mode on the string world-sheet in the primitive {\em
ansatz}. In what follows we will assume that the masses  $m_W$ and 
$m_a$ are of the same order of magnitude, i.e. $\tilde\lambda\sim\gs$.
In the logarithmic approximation  
 we can then
replace the logarithms in the first and the third
terms in Eq.~(\ref{sigmod}) by a single logarithm, say
ln$\,m_W/m_\gamma$. This assumption is by no means crucial.  

\subsection{Decays through tunneling (``bubbles of true vacua")}
\label{bubble}

To calculate the decay rate of the metastable string we may
forget for a while  about the microscopic theory 
(\ref{mod}) and turn our attention to the effective theory
(\ref{sigmod}). In this latter theory the string state
is nothing but ``the false vacuum state" 
at $\theta \approx 0$, see Fig. \ref{suf3},
while the ``no-string state" is the true vacuum at
$\theta \approx \pi/2$.

The metastable string decay occurs through the creation
of the monopole-like objects: at a certain $z=-z_0$ a magnetic charge  is
produced, accompanied by the production of an anticharge at $z=z_0$,
through tunneling. In the interval $-z_0 < z <z_0$ the magnetic flux tube is
eliminated. 

The corresponding process in the effective theory looks
as follows. In the initial moment of time the theory resides in the false
vacuum. Then it tunnels into the true one.
The tunneling creates an interval of true vacuum, which subsequently
experiences an unlimited classical expansion.

The quantitative description of the
false vacuum decay is well-developed within
the quasiclassical approximation
\cite{VKO,C}, which is fully applicable to the
2D sigma model (\ref{sigmod}). The applicability of the quasiclassical
approximation will become clear shortly. The parameter which regulates this
approximation is $V/v\gg 1$. 

Let us briefly review  the general procedure \cite{VKO,C}  
of calculating the probability of the false
vacuum decay (for  a comprehensive  review see \cite{V}). 
Details (as well as the final answer) 
slightly depend on the space-time dimension. Since our effective 
world-sheet theory (\ref{sigmod}) is (1+1)-dimensional, 
we will focus on this case.

\begin{figure}[htb]
\begin{center}
\includegraphics[width=11cm]{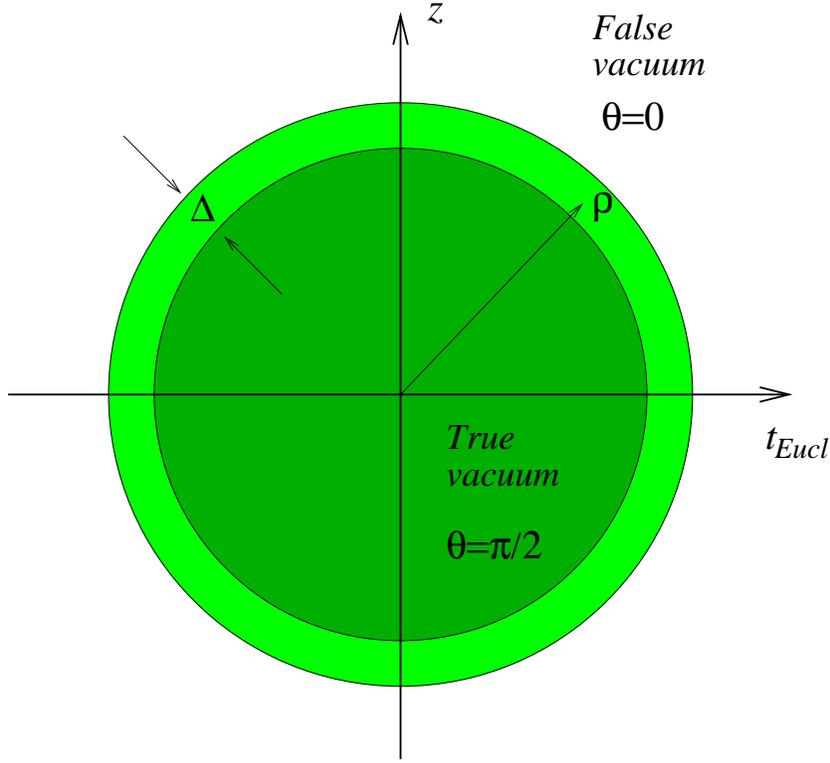}
\end{center}
\caption{
False vacuum decay through a bubble of true vacuum (in the Euclidean time).}
\label{xxx}
\end{figure}

An appropriate description of the tunneling probability 
implies a Euclidean rotation,
$$ t\to i\, t \,. $$
After the  Euclidean rotation,
the Euclidean action of the effective world-sheet theory
takes the form
\beqn
S_{\rm Eucl} = 2\pi \, \left( \ln {\frac{m_{W}}{m_{\gamma}}}\right)
\!\!\!\!\!\!\!\!\!\!
&&  \int  d^2
\sigma\,
\left\{\frac{2}{\gs}\left[(\dot\theta )^2 +(\partial_z\theta )^2
\right]+ \frac{V^2}2\sin^2{2\theta}  \right.
\nonumber\\[4mm]
&+&\left. 
\kappa \,  v^2 \cos^2{\theta} \right\}\,,
\label{ewst}
\\[4mm]
\kappa &\equiv& 
 \left( \ln {\frac{m_{q}}{m_{\gamma}}}\right)
\left( \ln {\frac{m_{W}}{m_{\gamma}}}\right)^{-1}
\, .
\label{ewstp}
\eeqn
In  this formulation the
false vacuum decay goes through the  creation of a bubble  of the 
true vacuum inside the false one, see Fig.~\ref{xxx}. 
The  bubble is a classical bounce solution  in the potential 
$V(\theta ) = -T^{\, \rm II}(\theta)$ which depends only
on the radial variable
$\sqrt{t^2+z^2}$.
In the thin wall approximation which is relevant to our problem
the concrete form of the bounce solution in the model (\ref{ewst})
is not  important. The only parameter we will need to know
is the tension  $\tau$ of the bubble surface.

The  bounce solution has a negative mode
associated with instability in the bubble size $\rho$. Integration over this
negative mode produces an imaginary part of the vacuum energy.
The latter determines the false vacuum  decay rate \cite{VKO,C} which thus is
proportional to 
\beq
\label{col}
\Gamma \sim \exp{\left(-S_{\rm bubble}\right)},
\eeq
where $S_{\rm bubble}$ is the classical action of the bounce.

In the problem at hand
the ratio of the critical size of the bubble to the bubble
surface thickness
is very large (it is regulated by the same parameter $V/v\gg1$).
Therefore, to calculate the bubble surface tension $\tau$ one can neglect its
curvature and consider a flat wall separating
two vacua --- one at $\theta \approx 0$
and another at $\theta \approx \pi /2$. Simultaneously we can (and should)
neglect the term $\kappa \,  v^2 \cos^2{\theta}$ in Eq.~(\ref{ewst})
which is responsible for the  non-degeneracy of
these two vacua. With the term $\kappa \,  v^2 \cos^2{\theta}$
switched off, the vacua at $\theta \approx 0$
and   $\theta \approx \pi /2$ become degenerate, and the 
flat wall perfectly stable.

The tension of the flat  wall is obtained from minimization
of the energy functional
\beq
\label{1d}
\tau =4\pi \left( \ln \frac{m_{W}}{m_{\gamma}}\right)\, 
\int \, d z \left\{
\frac1{g^2}\, \left( \partial_{z}\, \theta\right)^2 +
\frac{V^2}{4}\, \sin^2{2\theta}\right\}\, ,
\eeq
with the boundary conditions
\beq
\theta = 0\,\,\,\mbox{at}\,\,\, z\to -\infty\,,
\qquad \theta = \frac{\pi}{2} \,\,\,\mbox{at}\,\,\, z\to \infty\,.
\eeq
The solution of the minimization condition is a well-known
sine-Gordon soliton,
\beq
\theta (z) = \frac{1}{2}\, \left\{
{\rm arcsin}\left({\rm tanh} \, m_W z         \right) +\frac{\pi}{2}
\right\}\, .
\label{kink}
\eeq
The mass of the sine-Gordon kink gives the wall tension $\tau$,
\beq
\label{tau}
\tau = 4\, \pi \,\left(
\ln \frac{m_{W}}{m_{\gamma}} \right) \, \frac{V}{g}\, .
\eeq
The wall thickness is inversely proportional to the
$W$ boson mass,
\beq
\Delta\sim m_W^{-1}\,.
\label{wthi}
\eeq

The potential term in Eq.~(\ref{ewst}) presents  a huge barrier
under which the false vacuum must tunnel. At the same time, the energy density
difference   between the false and true   vacua
--- we denote it by $\E$ ---  is  small since it is determined by
the third term in Eq.~(\ref{ewst}),
\beq
\E\equiv T_{\rm ANO}^{\, \rm II}=2\pi \, v^2\,  \ln\, 
\left(  \frac{m_{q}}{m_{\gamma}}\right)\, .
\eeq
The action of the bubble on the tunneling trajectory is given by 
\cite{VKO}
\beq
\label{bact}
S_{\rm bubble}= 2\pi \rho \, \tau -\pi \rho^2\E\, .
\label{srho}
\eeq
The size of the critical bubble $\rho_*$
is determined by the extremum of the action,
\beq
\rho_* =  \frac{\tau}{\E}=2\left( \frac{V}{v}\right)^2 \frac{1}{gV\,\kappa}\,.
\label{crho}
\eeq
The ratio of the bubble radius to the wall thickness is indeed very large,
\beq
\frac{\rho_*}{\Delta}\sim \left( \frac{V}{v}\right)^2\gg 1\,,
\label{thinwall}
\eeq
which justifies the thin wall approximation. 

Substituting the critical size
from Eq.~(\ref{crho}) in Eq.~(\ref{srho}) we find the tunneling action,
$S_* = \pi {\tau^2}/{\E}$.
The probability of the false vacuum decay which is equal to the probability
(per unit time per unit length) of the ANO string breaking is
\beqn
\Gamma_{\rm breaking}^{\,\rm II} & \sim & e^{-S_*} =\exp\left(-
\pi\,\frac{\tau^2}{\E}\right) 
\nonumber\\[3mm]
&=& 
\exp \left\{ -\frac{8\pi^2}{\gs} \,\, \frac{V^2}{\kappa\, v^2}\,  
\left( \ln \, \frac{m_{W}}{m_{\gamma}} \right)\right\}
\,,
\label{gammaII}
\eeqn
where the parameter $\kappa$ is defined in Eq.~(\ref{ewstp}). 
 We conclude this section by
rewriting Eq.~(\ref{gammaII})
in terms of the ANO string tension (\ref{aten}),
\beq
\label{gammaten}
\Gamma_{\rm breaking}^{\,\rm II} \sim \exp\left\{ -\, \frac{16\pi^3}{\gs} 
\left( \ln\, \frac{m_{W}}{m_{\gamma}}\right)^2\, 
\frac{V^2}{T_{\rm ANO}^{\,\rm II}}\right\}\,.
\eeq
We will see in Sect.~\ref{five} that the decay rate of extreme type I
string, being  expressed in terms of the type I string tension, is given
by the very same formula.

\section{Extreme type I string}
\label{five}
\setcounter{equation}{0}

In this section we will consider strings in the limit of very small
quark masses. We still assume that the adjoint scalar mass
is much larger than all other scales, in other words
we impose the condition 
\beq
\label{I}
m_W\sim m_{a}\gg  m_{\gamma}\gg m_{q}\, .
\eeq
We will begin with a brief review of
the  stable ANO strings in this case, and then turn to decaying
strings. To avoid bulky expressions in what follows
it will be convenient to introduce a parameter $L$,
\beq
L= \ln\frac{m_\gamma}{m_q}\,,
\label{elll}
\eeq
and a function $L (\theta )$,
\beq
L  (\theta ) = \ln\, \frac{m_\gamma}{m_q\cos\theta}\,.
\label{ellltheta}
\eeq

\subsection{Type I ANO string}

Let us outline  the solution  \cite{Y99} 
for the ANO string in the Abelian Higgs model
(\ref{qed}) under the condition $m_{\gamma}\gg m_{q}$.

To the leading order in $L= \ln \, ( m_\gamma/m_q)$ the vortex
solution has the following structure. The
electromagnetic field is confined to a core with the radius $R_{g}$
which we will estimate momentarily.
The profile function for the gauge field is
given, approximately,  by an expression similar to Eq.~(\ref{f}),
\beq
\label{fI}
f=\left\{
\begin{array}{l}
1- \frac{r^2}{R_g^2}\, , \quad  r\ll R_g \, ,\\[3mm]
0, \qquad  \,\,\,\, \,\,\,\,      r\gg R_g \, .
\end{array}
\right.
\eeq
Moreover, the quark field is close to zero inside this core. On the other hand,
outside the core, the electromagnetic field is vanishingly
small. At intermediate distances
\begin{equation}
\label{logreg}
R_g\ \ll\ r\ \ll\ \frac1{m_q}
\end{equation}
the scalar field satisfies the free equation of motion, see
the first equation in (\ref{anoeq}), where the third and the last terms can be
ignored. Its
solution is as follows:
\begin{equation}
\label{qI}
q(r) =  v\left\{
1- \left(\ln\,\frac{1}{r m_q}\right)
\left(\ln\,\frac{1}{R_g m_q}
\right)^{-1} \right\}\,  .
\end{equation}
At large distances, $r\gg1/m_q$, the function  $q(r) $ approaches its
VEV $v$, the rate of approach is exponential,
$$q \to v\left(1+ C\, \exp(-m_q r)\right)\,.$$

Now,  substituting Eqs.~(\ref{fI}) and (\ref{qI}) in  Eq.~(\ref{anoten}) 
we get   the tension of the static type I  string as a function of $R_g$,
\begin{equation}
\label{Rten}
T_{\rm ANO}^{\, \rm I}\ =\ 2\pi\,
\left[\frac{\rm const}{g^2\,  R^2_{g}}   \, +\,  { v^2}\, 
\left(\ln\,\frac{1}{R_g m_q}
\right)^{-1}
\right] \,  .
\label{sundone}
\end{equation} 
The first term here comes from the first term in  Eq.~(\ref{anoten}))
which is concentrated inside the core.
The second term in Eq.~(\ref{sundone}) 
 comes from the
logarithmically large region (\ref{logreg}), where the quark  field is given
by Eq.~(\ref{qI}). 

Minimizing the right-hand side of Eq.~(\ref{Rten}) with respect to $R_g$ we 
determine $R_g$,
\begin{equation}
\label{Rg}
R^2_g\ \sim\
\frac{1}{m^2_\gamma}\,  L^2 =
\frac{1}{m^2_\gamma} \left( \ln \frac{m_\gamma}{m_q}\right)^2\, .
\end{equation}
This expression demonstrates  that  in the case at hand the size of
the string core
is logarithmically  larger than
$1/m_{\gamma}$, due to the presence of the  light scalar $q$.

Using this result for $R_g$ it is straightforward to evaluate the
tension of the extreme type I string. To this end we
plug  Eq.~(\ref{Rg})  back in   Eq.~(\ref{Rten}) and then obtain \cite{Y99}  
\begin{equation}
\label{yten}
T_{\rm ANO}^{\, \rm I}\ =\ {2\pi v^2}\, L^{-1}\,,
\end{equation}
where $L$ is defined in Eq.~(\ref{elll}).
The tension  $T_{\rm ANO}^{\, \rm I}$ 
is saturated, in the logarithmic approximation,
by the kinetic energy of the quark
field (the second term in  Eq.~(\ref{anoten}), the
``surface" energy). All other terms in Eq.~(\ref{anoten}), as well as corrections
to profile functions,  yield contributions
suppressed  by extra powers of $L^{-1}$.

\subsection{Decaying type I string}

We now turn to the derivation of an effective 2D world-sheet theory 
for the unstable mode $\theta (t,z)$ for the extreme type I strings.
Our first task is determination of the potential term.

The description of the decaying type I string we are going to use
runs parallel to that for type II, see Sects.~\ref{three} and \ref{four}.
The adjoint scalar 
$\varphi_{\theta}$ varies from its boundary value (\ref{0phibc}) 
to zero in a very
narrow core whose  size is of the order of
$1/m_{a}$. The profile function
$f_{\theta}$ for the electromagnetic field is concentrated inside a larger core,
of radius $R_{g} (\theta)$. 
Note that the parameter $R_g$ introduced above now becomes
a $\theta$ dependent function. 
Inside this ``electromagnetic" core the profile function
$f_{\theta}$ is approximately given by Eq.~(\ref{fI}) with
$R_{g}$ replaced by 
$R_{g} (\theta)$.
 The quark field is very small inside this core,  while
outside   it is given by Eq.~(\ref{qI}),
again with the replacement $R_{g}\to R_{g} (\theta)$. Assembling
all these elements together and substituting  in  Eq.~(\ref{ten}) we get
\beq
\label{RtenI}
T^{\rm I}(\theta)= 2\pi \, \left[
{\rm const}\,\, \frac{\cos^2{\theta}}{g^2 (R_{g}(\theta ))^2}\, +\, { v^2}\, 
\left(\ln\,\frac{1}{R_g m_q}
\right)^{-1}\,+ 
\left(  \ln\, \frac{m_{a}}{m_{\gamma}}\right) \,\frac{V^2}{2}\sin^2{2\theta}
\right]\, .
\eeq
Next, for each given $\theta$ one determines $R_{g} (\theta )$
by minimizing $T^{\rm I}(\theta)$ with respect to $R_g$.
In this way one  finds
\begin{equation}
\label{Rgth}
R_g (\theta)\ \sim\ \frac{\cos {\theta}}{m_\gamma}\, L\,,
\end{equation}
(cf. Eq.~(\ref{Rg})).
With this expression for $R_{g}(\theta) $ 
the tension  of the decaying type I string  versus $\theta$
(in the logarithmic approximation) takes the form
\beq
\label{tenI}
T^{\rm I}(\theta) = 
  \frac{2\pi \, v^2}{L (\theta )} +
2\pi  \,\left( \ln\, \frac{m_{a}}{m_{\gamma}}\right)
\,\frac{V^2}{2}\sin^2{2\theta}\,,
\eeq
where the function $L (\theta )$ is defined in Eq.~(\ref{ellltheta}).
The boundary values are as follows. At $\theta=0$
the potential term  $T^{\, \rm I}(\theta)$ is equal to the static  string tension 
(\ref{yten}).
At larger $\theta$ the potential    $T^{\rm I}(\theta)$ 
develops a very high barrier (at the maximum
$T^{\, \rm I} (\theta ) \sim V^2\,  \ln \, ( m_{a}/m_\gamma )$),  
and then it vanishes at $\theta=\pi /2$. Qualitatively, the behavior of
$T^{\, \rm I}(\theta)$ is perfectly the same as that depicted in Fig.~\ref{suf3}.
Equation  (\ref{tenI}) concludes our calculation of the
 potential term in the effective
world-sheet action for the $\theta$-mode of the type I string. 

Calculation of
 the kinetic term for  the type I string repeats the same steps we
made  in Sect.~\ref{43} for type II and thus
gives the  same result for the kinetic term of type I
as in Eq.~(\ref{sigmod}). It is easy to understand why: the
solution for the gauge  field is essentially the same for the two cases. 
Assembling
together the kinetic and potential terms we finally arrive at the 
following effective
world-sheet theory for the unstable mode of  the extreme type I string:
\beqn
S_{\rm str\,\, I} = 2\pi \, \int d^2 \sigma\,
\left\{\frac{2}{\gs}\left( \ln {\frac{m_{W}}{m_{\gamma}}}\right)
\,(\partial_{n}\theta)^2 - 
\frac{v^2}{L (\theta )}
- \left( 
\ln{\frac{m_{a}}{m_{\gamma}}}\right)
\,\frac{V^2}2\sin^2{2\theta}\right\}\, .
\label{sigmodI}
\eeqn

\vspace{2mm}

We are now ready to consider the  false vacuum decay in this sigma model.
As in Sect.~\ref{bubble}, the decay rate is given by the formula
\beq
\Gamma  \sim  \exp\left(-
\pi\,\frac{\tau^2}{\E}\right) \,,
\label{gamma54}
\eeq
where $\E$
is the difference between the energy densities in the false and true vacua,
$\E =T_{\rm ANO}^{\, \rm I}$, see Eq.~(\ref{yten}),
while $\tau$ is the tension of the flat
wall  separating the  two vacua. Remember, to calculate the latter we neglect
the term $v^2/L (\theta) $ in the potential energy.
Hence, the wall solution, as well as  $\tau$, are exactly the same as in 
Sect.~\ref{bubble}, see Eq.~(\ref{tau}), provided that $m_a \sim  m_W$, so that
the logarithms of $m_W/m_\gamma$ and $m_a/m_\gamma$ are the same. We 
accept this simplifying assumption.

 Substituting $\tau$ and $\E$ in 
Eq.~(\ref{gamma54}) we finally arrive at the 
decay rate (per unit length) of the extreme type I string,
\beq
\label{gammaI}
\Gamma^{\,\rm I}_{\rm breaking} \sim \exp\left\{ -\frac{8\pi^2}{\gs}\,
\frac{V^2}{v^2}\, 
\left( \ln\, \frac{m_{W}}{m_{\gamma}}\right)^2\, 
\left( \ln\, \frac{m_{q}}{m_{\gamma}}\right) 
\right\}\, .
\eeq
Needless to say that  if we express this decay rate in terms of the 
type I string tension
we get the same formula (\ref{gammaten}) as for the type II string,
\beq
\Gamma_{\rm breaking}^{\,\rm I} \sim \exp\left\{ -\, \frac{16\pi^3}{\gs} 
\left( \ln\, \frac{m_{W}}{m_{\gamma}}\right)^2\, 
\frac{V^2}{T_{\rm ANO}^{\,\rm I}}\right\}\,.
\label{tIano}
\eeq

\section{ Strings in the theory with two adjoint matter
fields}
\label{adjointsec}
\setcounter{equation}{0}

In this section we consider the decay problem for strings 
in the theory with two adjoint fields, see Eq.~(\ref{modprime}).
The light fields $\chi^{\pm}$ surviving in the ``macroscopic" 
  QED limit, see Eq.~(\ref{qedprime}), are written in  the matrix
form as follows:
\beq
\label{matchi}
\chi =\frac1{\sqrt{2}}\left(
\begin{array}{cc}
0 & \chi^{-}  \\
  \chi^{+} & 0 \\
 \end{array}\right).
\eeq
The ANO solution for the $Z_2$ string (\ref{anoprofprime}) corresponds to
the minimal winding of the scalar matrix,
\beq
\label{Z2chi}
\chi =\frac1{\sqrt{2}}\exp{\left(- i\, \frac{\tau_3}{2}\alpha\right)}\;\left(
\begin{array}{cc}
0 & 1  \\
  1 & 0 \\
 \end{array}\right)\;
\exp{\left( i\, \frac{\tau_3}{2}\alpha\right)}\;h_{1/2}(r)\, .
\eeq
In Fig. \ref{vstone} this trajectory corresponds to a semi-circle
running along the equator and connecting two points of intersection of the equator
with the 1-st axis. As we have already mentioned this string is stable --- one cannot
 unwind it. 

However,   strings with multiple winding numbers are metastable.
Let us use the method developed in the previous sections to calculate
the decay rate of the string with the winding number $n=2$.
The winding of the scalar field  $\chi$ in 
Eq.~(\ref{anoprofpr})   in the
matrix notation takes the form
\beq
\label{n2chi}
\chi =\frac1{\sqrt{2}}\exp{\left(-i\, \tau_3\alpha\right)}\;\left(
\begin{array}{cc}
0 & 1  \\
  1 & 0 \\
 \end{array}\right)\;
\exp{\left(i\, \tau_3\alpha\right)}\; h(r)\, .
\eeq
Here the scalar profile function satisfies the 
boundary conditions
\beq
\label{hbc}
h(0)=0,\; h(\infty)=v.
\eeq

To unwind this string we can use the same ``unwinding'' matrix
(\ref{matrix}) which we used in the theory with the  fundamental scalar.
The ``unwinding'' {\em ansatz} takes the form
\begin{eqnarray}
\chi(x) &=&
 \frac1{\sqrt{2}}U\,\tau_1\,U^{-1}\, h_{\theta} (r)\, ,
 \nonumber\\[2mm]
A_0 &\equiv& 0\,, \qquad A_3  \equiv  0\,, \nonumber\\[2mm]
 A_j(x)   &=& i \, U\partial_{j}\, U^{-1} \, [1-f_{\theta}(r)]\, , \qquad j=1,2\,,
 \nonumber\\[3mm]
\phi &=& V\,U\, \frac{\tau_{3}}{2}\, U^{-1} +\Delta\phi \, ,
\label{anzad}
\end{eqnarray}
where $\Delta \phi$ is given by (\ref{deltaphi}). Substituting this 
{\em ansatz} in  the action (\ref{modprime}) we get
\begin{eqnarray}
T(\theta)
&=&\!\!\!
2\pi \int rdr\,\left\{
\frac{2}{g^2}\, \frac{f_{\theta}'^2}{r^2}\,
\cos^2{\theta}
 + h_{\theta}'^2
+4\frac{f_{\theta}^2}{r^2}
\,
h_{\theta}^2 \, \cos^2{\theta}\left(1-\frac12 \sin^2{\theta}\right)
\right.
\nonumber\\[3mm]
&+&\!\!\!
\frac{1}{2}\varphi_{\theta} '^2 +\frac1{2r^2}\left[
\varphi_{\theta}(\cos{2\theta}-
2f_{\theta}\cos^2{\theta})+Vf_{\theta}\sin{2\theta}\right]^2
\nonumber\\[3mm]
&+&\!\!\!
\left.
\beta (h_{\theta}^2-v^2)^2+
\alpha\left[
\varphi_{\theta}^2
-2V\varphi_{\theta}\sin{2\theta}\right]^2 
+\frac{\gamma}{4}
\varphi_{\theta}^2 h_{\theta}^2\cos^2{2\theta}\right\}\, .
\label{tenad}
\end{eqnarray}
Comparing this result with Eq.~(\ref{ten}), we see that the only terms
which are modified   are the ones associated with the light
scalar $\chi$. In particular, the terms coming from the heavy
scalar $\phi$ , responsible for the huge potential barrier,  stay intact.

To proceed,  let us consider the case of the extreme type II
string assuming that the light scalar mass $m_{\chi}$
is much larger than the photon mass, $m_{\chi}\gg m_{\gamma}$.
Using essentially the same step-function model  for the profile functions
in (\ref{anzad}) as in Sects. \ref{arcone} and \ref{43},  we finally arrive at the 
following effective  sigma model for the unstable mode on the 
string world sheet:
\beqn
S_{\rm str\,\, II} &=& 2\pi \, \int d^2 \sigma\,
\left\{\frac{2}{\gs}\left( \ln {\frac{m_{W}}{m_{\gamma}}}\right)
\,(\partial_{n}\theta)^2 \right.
\nonumber\\[3mm]
&-&\left. \! \! \! \! 
4\left( \ln{\frac{m_{q}}{m_{\gamma}}}\right) \, v^2 \cos^2{\theta}
\left(1-\frac12 \sin^2{\theta}\right)
-\left( 
\ln{\frac{m_{a}}{m_{\gamma}}}\right)
\,\frac{V^2}2\sin^2{2\theta}\right\}\, .
\label{sigmodad}
\eeqn
At $\theta=0$ the potential in this sigma model
 reduces to the tension of $n=2$ ANO string,
\beq
T_{\rm ANO}^{\rm II,\,n=2}=2\pi v^2\,4  \ln\, \frac{m_{q}}{m_{\gamma}}\,.
\label{atenad}
\eeq
The extra factor 4 here, as compared to Eq.~(\ref{aten}), is due to the $n^2$ 
dependence of the extreme type II string tension on the winding number $n$.

Now, to calculate the decay rate of this string we follow the same steps as in 
Sect. \ref{bubble}. Note that the tension of the domain wall is still given 
by Eq.~(\ref{1d}) and the only modification is due to the expression for
$\E$ which is now given by the $n=2$ ANO string tension
(\ref{atenad}), $\E =T_{\rm ANO}^{\rm II,\,n=2}$. It is clear that, being
expressed in terms of the string tension, the decay rate is given
by the same expression (\ref{gammaten}),
\beq
\Gamma_{\rm breaking}^{\,\rm II,} \sim \exp\left\{ -\, \frac{16\pi^3}{\gs} 
\left( \ln\, \frac{m_{W}}{m_{\gamma}}\right)^2\, 
\frac{V^2}{T_{\rm ANO}^{\,\rm II,\, n=2}}\right\}\,.
\label{gammaad}
\eeq
 
We can calculate, with the same ease,  the decay rate of the extreme type I string
in the theory with two adjoint scalars starting from Eq.~(\ref{tenad})
and following the same procedure as in Sect.~\ref{five}. Clearly, the result for the
decay rate, when expressed in terms of the ANO string tension 
$T_{\rm ANO}^{\,\rm I,\, n=2}$, is given
by the same formula (\ref{gammaad}).

\section{Lessons}
\label{doptwo}
\setcounter{equation}{0}

We pause here to summarize  what we have learned
from the calculation above and to abstract general features.

\begin{figure}[htb]
\begin{center}
\includegraphics[width=9cm]{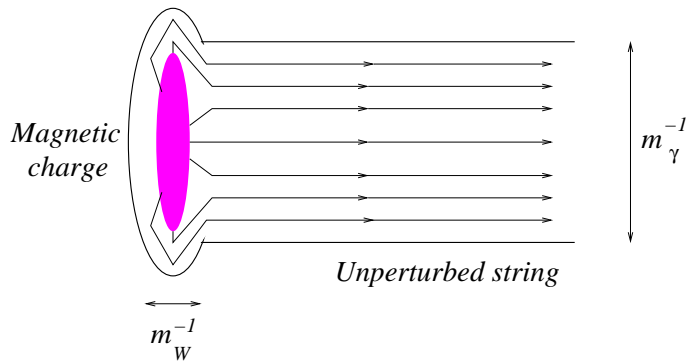}
\end{center}
\caption{
The right-hand half of the broken  string (in the {\em ansatz}
(\ref{anz})). One unit of the magnetic charge is produced in the shaded area.
Arrows mark the magnetic field flux.}
\label{vsttwo}
\end{figure}

The kink (\ref{kink}) represents a lump of energy, a bulge
at the end of the broken string associated with the production
of one unit of the magnetic charge, see Fig. \ref{vsttwo}.

The kink mass $\tau$ (see Eq. (\ref{tau})) is the mass of the
bulge carrying one unit of the magnetic charge.
Keeping in mind that $M_M\sim V/g$ one may ask why
$\tau $ is logarithmically larger than $M_M$.

The answer to this question is quite obvious. While the   {\em ansatz}
(\ref{anz}) does describe the production of the magnetic charge at the end of the
broken string, it is a highly excited monopole-like state rather than the 
't~Hooft-Polyakov monopole.
Indeed, the longitudinal dimension of the bulge is $\sim m_W^{-1}$, a typical
size of the monopole core. At the same time its transverse dimension
(in the plane perpendicular to the string action) is of order 
$  m_\gamma^{-1}$. This is much larger than the monopole core size.
The stretching of the core in the perpendicular direction is the reason why
this lump is logarithmically heavier than the 't Hooft-Polyakov monopole.
This is an inevitable consequences of the fact that the 
 {\em ansatz}
(\ref{anz}) contains a single profile function $f_\theta (r)$
which governs the behavior of both, the photon and the $W$ boson fields.

By introducing two distinct profile functions one can readily eliminate the
shortcoming in  the 
 {\em ansatz}
(\ref{anz}).  We  do it in Appendix. However, even before
particular improvements are done we want to note that   
 sufficient experience is already accumulated 
to enable us  to develop a proper effective description
of the tunneling process.

Indeed, in actuality the end-point domain of the broken string is
roughly a hemisphere with the radius $\sim  m_\gamma^{-1}$.
The core which emanates the magnetic flux has dimension of order
of $ m_W^{-1}$ in both transverse and longitudinal directions.
In fact, the core is practically unperturbed 't~Hooft-Polyakov monopole.
This is due to the fact that at distances of order $ m_W^{-1}$
the effect of the (magnetic charge) confinement is negligible, it comes into
play only at distances $\sim  m_\gamma^{-1}$. Thus, the mass of the
end-point bulge in a ``good" {\em ansatz} must be $\tau = M_M
+O(v/g)$. The $O(v/g)$ correction reflects the distortion of the
't~Hooft-Polyakov monopole at distances $\sim  m_\gamma^{-1}$, and can be
neglected compared to $M_M$. A potential of the type depicted
in Fig. \ref{suf3} will emerge leading to the tunneling problem
described by the bubble action (see Appendix for details),
\beq
S_{\rm bubble} = 2\pi\,\rho M_M -\pi\rho^2 T_{\,\rm ANO}\,.
\label{sunone}
\eeq
The only distinction with the consideration carried out in Sect.~\ref{bubble}
is that in the ``good" {\em ansatz} the ratio of the
bubble radius to its thickness
\beq
\frac{\rho_*}{\Delta}\sim \left( \frac{V}{v}\right) 
\label{suntwo}
\eeq
(cf. Eq.~(\ref{thinwall})). We loose one power of $V/v$,
but the ratio ${\rho_*}/{\Delta}$ is still large, and the thin wall approximation
justified.

The action (\ref{sunone}) immediately leads to the
string decay rate given in Eq.~(\ref{ura}). The effective approach based on
Eq.~(\ref{sunone}) is exactly the one of Vilenkin~\cite{Vil} and Preskill and
Vilenkin~\cite{PV}.

In view of a general nature of this conclusion, let
us comment on the 
  hierarchy of parameters we deal with.
We want to separate elements of this hierarchy which are
absolutely essential for our consideration from those which bear a technical
character.

Our consideration is quasiclassical. It is valid only at weak coupling.
This requires Eq.~(\ref{1one}) to be valid.
We cannot sacrifice this condition. 

Moreover, the two-stage nature of the gauge  symmetry breaking,
Eq.~(\ref{2two}), is crucial too. Among other things,
it guarantees that the thin wall approximation is justified.
It is unclear   whether lifting
the  condition (\ref{2two}) one can still develop an analytic description of the
tunneling process. We will not try to lift the constraint (\ref{2two}).

On the other hand, the relation between
$m_q$ and $m_\gamma$ is clearly of a technical nature.
In the two extreme cases, when 
$m_q/m_\gamma$ is either very large or very small
one can calculate  ANO string tension analytically.
One can pose a question what happens at arbitrary values
of $m_q/m_\gamma$. Calculation of the ANO string tension
in this case certainly requires numerical computations.
However, this calculation can be carried out
entirely in the macroscopic low-energy U(1) theory, with no reference
to the microscopic non-Abelian theory.
The procedure is well-developed; the function $F(m_q/m_\gamma )$
which parametrizes the ANO string tension in the general case, 
\beq
T_{\rm ANO} = 2\pi v^2 \, F\left(\frac{m_q}{m_\gamma}\right)\,,
\eeq
is known in the literature \cite{BogoVa}. 

Irrespective of the value of $m_q/m_\gamma$,
the energy difference in the false and true vacua is $\E = T_{\rm ANO}$.
Moreover, the expression (\ref{ura}) for the decay rate is 
universally valid as long as $\rho_*/\Delta \gg 1$, i.e.
$V/v\gg 1$. The flat wall tension $\tau$ is calculated
in the approximation which neglects
the quark field  contributions   altogether ---
the barrier is  determined only by  terms proportional to $V^2$.
No matter what the particular {\em ansatz} is,
it must yield $\tau = M_M$ modulo possible small correction
$\sim v/g$. This condition can be
viewed as a test of the ``goodness" of the {\em ansatz}.

Therefore, if one
neglects all terms suppressed by powers
of $v/V$ 
one {\em inevitably} arrives at Eq.~(\ref{sunone}), irrespective of the ratio
$m_q/m_\gamma $. The result (\ref{ura}) for  for the string decay rate
$\Gamma_{\rm breaking}$ ensues. It is valid for  arbitrary value  of the ratio
$m_q/m_\gamma$. 

Of particular interest are examples emerging in the
supersymmetric setting.
For instance, for the  BPS string 
$T_{\rm ANO}=2\pi v^2$.
Moreover, the BPS bound for the monopole mass $M_{M}$ is
\beq
\label{Mm}
M_M=\frac{4\sqrt{2}\pi}{g}\,V\, .
\eeq
Then  Eq.~(\ref{ura})  predicts the following    decay rate of this string:
\beq
\label{gammabps}
\Gamma_{\rm BPS} \sim \exp
\left\{ -\frac{8\pi^2}{\gs}\, \frac{2\, V^2}{v^2}\,
\right\}\, .
\eeq
Let us recall that the Abelian BPS strings embedded in non-Abelian gauge theories
appear, say, in the charge vacua of the Seiberg-Witten theory with matter
\cite{HSZ,FG,VY,MY}.

\section{Comparison with Schwinger's expression}
\label{six}
\setcounter{equation}{0}

The decay of the string goes through  breaking of the string in 
pieces through a monopole-antimonopole pair production. 
From this standpoint,  
the bubbles of the ``true vacuum'' of the effective 2D sigma model
inside the ``false'' one are, in fact,   domains where the string is
broken by monopole-antimonopole pairs. The decay rate is
exponentially small for large monopole masses; the exponent is
determined by the ratio   $M_M^2/T_{\rm ANO}$.

\begin{figure}[htb]
\begin{center}
\includegraphics[width=6cm]{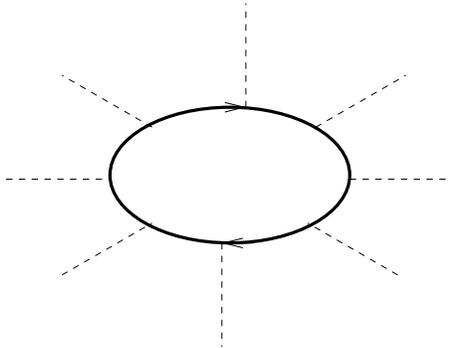}
\end{center}
\caption{
Propagation of the electric charge in the background electric field.
The charged particle loop is denoted by a thick line,
the background field by dashed lines.}
\label{yyy}
\end{figure}

Given this interpretation it is instructive to compare  
Eq.~(\ref{ura}) with the famous Schwinger's formula \cite{Schw,AfMa}
for the electron-positron pair production in the constant electric field,
\beq
\Gamma_{e^+e^-} \sim \exp\left(-\pi\,\frac{m^2}{E}
\right)\,,
\label{tuesdayone}
\eeq
where $m$ is the electron mass, and $E$ is the electric field \footnote{
Extensions of the Schwinger formula in string theory were recently 
discussed in Refs.~\cite{qu,ququ}.}.
Note that with our normalization, see Eq.~(\ref{mod}), 
the coupling constant is included in $E$. The probability (\ref{tuesdayone})
can be obtained as the imaginary part of the one-loop graph presented in
Fig.~\ref{yyy}.

Dualizing Schwinger's expression we can 
try to use it  for evaluating  the probability of the monopole-antimonopole pair
creation in the  magnetic field existing in the core of the ANO string. 
Of course, we have to assume that this field is constant
on the scale of the monopole-antimonopole separation (we hasten to add that
this is a wrong assumption).

It is not difficult to get from Eq.~(\ref{tuesdayone})
a dualized Schwinger formula for the magnetic monopole
pair production in the homogeneous
magnetic field. Indeed, with our normalization the duality transformation
 reads $E\leftrightarrow g^{-2} B$ where $B$ is the magnetic field.
Then the dualized Schwinger formula takes the form
\beq
\label{schw}
\Gamma_{M\bar M}\sim \exp{\left(-\pi\,\, \frac{M_M^2}{B\, g^{-2}}\right)}\,.
\eeq
The magnetic field in the ANO string can be readily
estimated from   its   flux,
\beq
B\sim \frac{2\pi }{\rm area} \sim   m_\gamma^2 =  \frac{1}{2}\,  g^2\,v^2\,.
\label{tuesdaythree}
\eeq
 Combining Eqs. (\ref{schw})
and (\ref{tuesdaythree}) one obtains 
\beq
\Gamma_{M\bar M}\sim \exp{\left(-\pi\, C\, \frac{M_M^2}{v^2
}\right)}\,,
\label{sc}
\eeq
where $C$ is a numerical coefficient of order 1.

\begin{figure}[htb]
\begin{center}
\includegraphics[width=6cm]{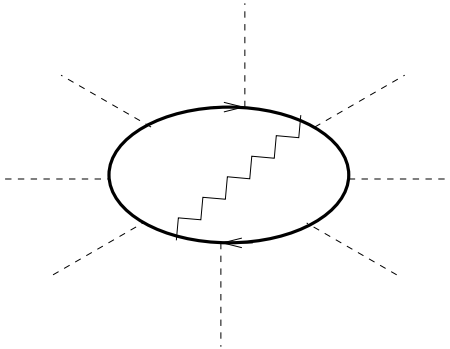}
\end{center}
\caption{Propagation of the monopole
 in the background  magnetic field.
The monopole loop is denoted by a thick line. $\gamma$ quanta
(denoted by a
zigzagy line) are strongly coupled to the monopoles, and 
exchange of the $\gamma$ quanta must be taken into account.}
\label{zzz}
\end{figure}

One might ask whether dualizing Schwinger's formula is a good idea. Indeed,
  Schwinger's formula assumes the smallness of the gauge coupling constant
$g^2$. The coupling of the magnetic monopoles $\sim g^{-2}$
is then strong. Therefore, the $\gamma$ quanta exchanges, as in Fig.~\ref{zzz},
might drastically change the result. 

This effect was studied
in Ref.~\cite{AfMa}, where it was found that the extra term in the exponent
due to the gamma quanta exchanges is of order $1/g^2$.
This  is easy to understand. Indeed, the corresponding Coulomb energy
is of order $1/(\rho_*\, g^2)$ which generates a correction in the bubble action of 
the order
of $1/g^2$. It can be safely neglected because  $M_M^2/v^2 \gg 1/g^2$.

Comparing this Schwinger-formula-based expectation 
with our result (\ref{ura}), and keeping in mind that
$T_{\rm ANO} = 2\pi v^2$ modulo a logarithm of
$m_q/m_\gamma$, 
we see that the powers of the mass scales and of the coupling $g^2$
in the two  exponents perfectly match each other.
However, the logarithmic factors 
$\left( \ln\,  \frac{m_{q}}{m_{\gamma}}\right)^\kappa$
are missing in (\ref{sc}).
These will not (and need not) match, and neither  a numerical factor
in front of $M_M^2/v^2$. The reason for this  is that Schwinger's
formula and our calculation refers to physically different phases of the 
theory. Schwinger's
formula (\ref{schw}) describes monopole-antimonopole production in the
external magnetic field in the Coulomb phase while our result
(\ref{ura}) refers to the breaking of the ANO string by the 
monopole-antimonopole production in the confinement (for monopoles)
phase of the theory.

\section{Conclusions}
\label{seven}

In this paper we calculated the decay rate of an Abelian flux tube
embedded into a non-Abelian theory. 
 We focussed on two 
simplest examples where the phenomenon does occur:
 non-supersymmetric SU(2) gauge theory with the adjoint and
fundamental scalars and the same theory with two adjoint scalars.
In the first example all ANO strings are metastable. In the second
one the string with minimal winding number is stable ($Z_2$-string),
while strings with multiple winding numbers are metastable.

 The pattern of the SU(2) symmetry breaking is two-stage,
$$
{\rm SU(2)} \stackrel{V}{\longrightarrow} {\rm U(1)}
\stackrel{v}{\longrightarrow} {\rm nothing}\,,\qquad V\gg v\,.
$$
As was expected, the decay rate is exponentially small.
The suppressing exponent is proportional to
 the ratio of monopole mass squared
to the string tension. The interpretation of this result is that the string
gets broken into pieces by the monopole-anti-monopole production.
Note that the monopole-anti-monopole pair production in the given context
by no means implies that superconductivity is lost.

Although we considered a particular model with  metastable strings, we
 believe
that the final  answer is rather general and can be qualitatively applied to
any metastable Abelian string embedded in a non-Abelian theory.
In particular, as we mentioned in Sect.~\ref{one},  the reduction of
string multiplicity from $Z^{N-1}$ down to $Z_N$ in the strong coupling
vacua of the Seiberg-Witten theory is due to a similar mechanism.  
In this case we deal with electric strings (which arise  due
to the  monopole /dyon   condensation), so that the metastable strings must
broken by the $W$ boson pair production (rather than the monopole 
pair production which takes place in the 
 magnetic flux tubes).

Unwinding {\em ans\"{a}tze} of the type presented in Eq.~(\ref{anz})
can be used in other similar problems, for instance, for studying the
metastability of the appropriately embedded semilocal strings
(cf. \cite{PV}, for a review of the semilocal strings see Ref.~\cite{SLS}).

Finally, it is worth noting that the calculation of the string decay rate
presented here can be viewed as a calculation of an open string 
coupling constant in the effective string theory of ANO string.

\section*{Acknowledgments}

We are grateful to G. Dunne for a discussion
which stimulated our search for   analytic solution
of the string decay rate problem. 
We would like to thank A.~Gorsky, M.~Kneipp,  A.~Vainshtein and M.~Voloshin
for valuable discussions and comments, and I.~Khriplovich for providing
useful references. Two crucial references were pointed out to us 
by a Phys. Rev. D referee.
A. Y. would like to thank the Theoretical Physics
Institute, University of Minnesota, where this
work was carried out,  for hospitality and support.

This work 
is supported in part by the DOE grant 
DE-FG02-94ER408 and  CRDF grant  CRDF RP1--2108. A.Y. is also
supported by 
Russian Foundation for Basic Research grant   02-02-17115,   
INTAS grant   2000-334 ,  and  Support of Scientific
 Schools grant  00-15-96611.

\section*{Appendix. Improving the primitive unwinding  \\
{\em ansatz}}
\label{dopthree}
\renewcommand{\theequation}{A.\arabic{equation}}
\setcounter{equation}{0}

As we explained in Sect.~\ref{doptwo} our {\em ansatz} (\ref{anz}) gives  
a monopole in a highly exited state at the end of the broken string.
Its mass $\tau$ is much larger then the monopole mass due to the 
logarithmic factor in (\ref{tau}). The reason is that we used
the same profile function $f_{\theta}$ for both $a=3$ and $a=1,2$
components of the gauge potential in (\ref{anz}). We used the model
(\ref{f}) for this profile function which ensures that all components
of the gauge field are nonvanishing  inside the region of  size
 $\sim 1/m_{\gamma}$ in the (1,2)-plane. 
However, it is clear that the  $a=1,2$ components of the gauge field are
very heavy (W bosons) and should be spread over a much smaller
region, of size $\sim 1/m_{W}$. Now, let us modify the {\em ansatz}
(\ref{anz}) to take  this circumstance into account.

First, let us explicitly write down the gauge field given by the
{\em ansatz} (\ref{anz}) in the singular gauge, see Eq.~(\ref{singan}),
\beq
A_{i}= - \, 
\varepsilon_{ij}\,
\frac{x_j}{r^2}\, 
\left[ 2\cos^2{2\theta}\,\, \frac{\tau_3}{2}-
\sin{2\theta}\left(\frac{\tau_1}{2}\sin{\alpha}
+\frac{\tau_2}{2}\cos{\alpha}\right)\right]\,f_{\theta}(r)\,  .
\label{oldA}
\eeq
To modify it, we introduce two different profile functions,
in front of $\tau_3$ and $\tau_{1,2}$ matrices, as follows:
\beq
A_{i}= - \, 
\varepsilon_{ij}\,
\frac{x_j}{r^2}\, 
\left[ 2\cos^2{2\theta}\,\, \frac{\tau_3}{2}\,f^{\gamma}_{\theta}(r)-
\sin{2\theta}\left(\frac{\tau_1}{2}\sin{\alpha}
+\frac{\tau_2}{2}\cos{\alpha}\right)\,f^{W}_{\theta}(r)\right] .
\label{newA}
\eeq
We assume that $f^{W}$ is concentrated inside a very small region
of size $\sim 1/m_{W}$, whereas $f^{\gamma}$ is spread over a much
larger domain of size $\sim 1/m_{\gamma}$. The expressions for the 
adjoint and the  fundamental
 scalars are still given by Eq.~(\ref{singan}).

Substituting this into the action (\ref{mod}) we end up with  
\begin{eqnarray}
T(\theta)
&=&\!\!\!
2\pi \int rdr\,\left\{
\frac{2}{g^2}\, \frac{\cos^2{\theta}}{r^2}\,
\left[\cos^2{\theta}(f_{\theta}^{\gamma})'\,^2 
+\sin^2{\theta}(f_{\theta}^{W})'\,^2\right]
 +q_{\theta}\,'\,^2
\right.
\nonumber\\[3mm]
&+& \frac{q_{\theta}^2}{r^2}
\,
  \cos^2{\theta}
\left(\cos^2{\theta}f_{\theta}^{\gamma} +\sin^2{\theta}f_{\theta}^{W}\right)
\nonumber\\[3mm]
&+&\!\!\!
\frac{1}{2}\varphi_{\theta} '\,^2 +\frac1{2r^2}\left[
\varphi_{\theta}\left(\cos{2\theta}\,( 1-f_{\theta}^{\gamma})-
(\cos^2{2\theta}f_{\theta}^{\gamma} +\sin^2{2\theta}f_{\theta}^{W})\,
\right)+Vf^{W}_{\theta}\sin{2\theta}\right]^2
\nonumber\\[3mm]
&+&\!\!\!
\left.
\lambda \left(q_{\theta}^2-v^2\right)^2+
\tilde{\lambda}\left[
\varphi_{\theta}^2
-2V\varphi_{\theta}\sin{2\theta}\right]^2 
+\frac{\gamma}{4}
\varphi_{\theta}^2q_{\theta}^2\right\}\, .
\label{tenmod}
\end{eqnarray}

We see that now  the term $Vf^{W}_{\theta}\sin{2\theta}$ 
in the square brackets contains $f^{W}$
rather than $f^{\gamma}$ (cf. Eq.~(\ref{ten})). This means that its contribution
in the 
integral will be saturated in a  a small region, of   size $\sim 1/m_W$,  and 
will not lead to 
 large logarithms. Thus, the height of the barrier now is
$V^2$ instead of $V^2\,\ln(m_a/m_{\gamma})$. 
It should be added, however, that in the absence of large
logarithms  we cannot
obtain the barrier profile using analytic calculations. We need 
computer simulations to minimize the   tension (\ref{tenmod}) and find
the profile functions. This goes beyond the scope of this paper.

Still,  some  simple examples of the  profile functions   we have analyzed
show  that we can avoid getting large logarithms both in
the barrier height and in the kinetic term for the $\theta$ field
in the effective 2D sigma model for the unstable string mode.
This leads us to the following representation for the domain wall
tension $\tau$
\beq
\label{1dmod}
\tau ={\rm const}\;  
\int \, d z \left\{
\frac1{g^2}\, \left( \partial_{z}\, \theta\right)^2 +
\frac{V^2}{4}\, \sin^2{2\theta}\right\}\, ,
\label{arctwo}
\eeq
instead of the one in Eq.~(\ref{1d}). Here, the constant in front
of this action can be fixed by means of a numerical minimization procedure
in (\ref{tenmod}). Note, that the relative coefficient between
the kinetic and the potential terms in Eq.~(\ref{arctwo})  is fixed by the 
requirement
that the mass of  the $\theta$ field should coincide  with that of the 
$W$ boson.

Calculating the domain wall tension in the theory (\ref{1dmod}) we get
\beq
\tau\,={\rm const}\; \frac{V}{g}\, .
\label{taumod}
\eeq
In order to prove that actually 
\beq
\tau=M_M
\label{taumon}
\eeq
 we need numerical
simulations. Still, if we accept on physical grounds that this is
the case, then the bubble action in the 2D sigma model   given by
Eq.~(\ref{sunone}) ensues. This leads us to the final formula (\ref{ura}) for 
the decay rate of the metastable string.


\newpage


\begin{thebibliography}{99}

\bibitem{SW1}
N.~Seiberg and E.~Witten, Nucl. Phys. {\bf B426}, 19  (1994) 
[hep-th/9407087].

\bibitem{SW2}
N.~Seiberg and E.~Witten, Nucl. Phys. {\bf B431}, 484  (1994),
[hep-th/9408099].

\bibitem{A}
A.~Abrikosov,  ZhETF {\bf32}, 1442  (1957) [Eng. transl. Sov.~Phys. JETP,
{\bf 5}, 1174 (1957); reprinted in {\em Solitons and Particles}, Eds. C. Rebbi 
and G.~Soliani,
(World Scientific, Singapore, 1984), p. 356].

\bibitem{NO}
H.~B.~Nielsen and P.~Olesen,
Nucl.\ Phys.\ B {\bf 61}, 45 (1973).
[Reprinted in {\em Solitons and Particles}, Eds. C. Rebbi and G. Soliani,
(World Scientific, Singapore, 1984), p. 365].

\bibitem{B}
E.~B.~Bogomolny,
Yad.\ Fiz.\  {\bf 24}, 861 (1976) [Engl. transl. Sov.\ J.\ Nucl.\ Phys.\  {\bf 24}, 449 (1976);
 reprinted in {\em Solitons and Particles}, Eds. C. Rebbi and G.~Soliani,
(World Scientific, Singapore, 1984), p. 389].

\bibitem{deV}
H.J.~de Vega, Phys. Rev. {\bf D18} (1978) 2932.

\bibitem{VS}
H.J.~de Vega and F.A.~Shaposnik, Phys. Rev. Lett.{\bf 56} (1986) 2564;
Phys. Rev. {\bf D34} (1986) 3206.

\bibitem{HV}
J.~Heo and T.~Vachaspati, Phys. Rev. {\bf D58} (1998) 065011
[hep-th/9801455].

\bibitem{SS}
F.A.~Shaposnik and P.~Suranyi, Phys. Rev. {\bf D62} (2000) 125002
[hep-th/0005109].

\bibitem{KB}
 M.~Kneipp and P.~Brockill, 
Phys. Rev. {\bf D64} (2001) 125012 [hep-th/0104171].

\bibitem{KS} 
K.~Konishi and L.~Spanu, {\em ``Non-Abelian vortex and
confinement''}, hep-th/0106175.


\bibitem{DS}
M.~Douglas and S.~Shenker, Nucl. Phys. {\bf B447}, 271 (1995) 
 [hep-th/9503163].
 
\bibitem{S}
M.~Strassler, Prog. Theor. Phys. Suppl. {\bf 131},   439 (1998) 
[hep-th/9803009].


\bibitem{HSZ}
A.~Hanany, M.~Strassler and A.~Zaffaroni,
Nucl. Phys. {\bf B513}, 87 (1998) [hep-th/9707244].

\bibitem{Y99}
A.~Yung,
Nucl.\ Phys.\  {\bf B562}, 191 (1999) [hep-th/9906243].

\bibitem{Y01}
A.~Yung,
Nucl. Phys. {\bf B626}, 207 (2002) [hep-th/0103222].

\bibitem{MY}
A.~Marshakov and A.~Yung,
{\em Non-Abelian confinement via Abelian flux tubes 
in softly broken ${\cal N} = 2$  SUSY QCD},
hep-th/0202172.

\bibitem{thooft}
G.~'t Hooft,
Nucl.\ Phys.\ B {\bf 79}, 276 (1974).

\bibitem{polyakov}
A.~M.~Polyakov,
Pisma Zh.\ Eksp.\ Teor.\ Fiz.\  {\bf 20}, 430 (1974)
[Engl. transl. JETP Lett.\  {\bf 20}, 194 (1974),
reprinted in {\em Solitons and Particles}, Eds. C. Rebbi and G. Soliani,
(World Scientific, Singapore, 1984), p. 522].

\bibitem{Schw}
F. Sauter, Z. Phys. {\bf 69}, 742 (1931);\\
W. Heisenberg and  H. Euler, Z. Phys. {\bf 98}, 714 (1936);\\
J. Schwinger, Phys. Rev. {\bf 82},   664 (1951).

\bibitem{AfMa}
I.~K.~Affleck and N.~S.~Manton,
Nucl.\ Phys.\ B {\bf 194}, 38 (1982).

\bibitem{Vil}
A.~Vilenkin,
Nucl.\ Phys.\ B {\bf 196}, 240 (1982).

\bibitem{PV}
J.~Preskill and A.~Vilenkin,
Phys. Rev. {\bf D47}, 2324 (1993).

\bibitem{VKO}
M.~B.~Voloshin, I.~Y.~Kobzarev, and L.~B.~Okun,
Yad.\ Fiz.\  {\bf 20}, 1229 (1975)
[Sov.\ J.\ Nucl.\ Phys.\  {\bf 20}, 644 (1975)].

\bibitem{C}
S.~R.~Coleman,
Phys.\ Rev.\ D {\bf 15}, 2929 (1977);
(E)  D {\bf 16}, 1248 (1977) [Reprinted in {\em
The Early Universe}, Eds. E.W. Kolb and M.S. Turner, (Addison-Wesley, 1990), p. 483].


\bibitem{HS}
Z.~Hlou\v{s}ek and D.~Spector,
Nucl.\ Phys.\ B {\bf 370}, 143 (1992);\\
J.~Edelstein, C.~Nun\~ez and F.~Schaposnik,
Phys.\ Lett.\ B {\bf 329}, 39 (1994)
[hep-th/9311055].

\bibitem{DDT}
S.~C.~Davis, A.~C.~Davis and M.~Trodden,
Phys.\ Lett.\ B {\bf 405}, 257 (1997)
[hep-ph/9702360].

\bibitem{GS} 
A.~Gorsky and M.~A.~Shifman,
Phys.\ Rev.\ D {\bf 61}, 085001 (2000)
[hep-th/9909015].

\bibitem{FG}
 W.~Fuertes  and J.~Guilarte, 
Phys. Lett. {\bf B437},  82 (1998) [hep-th/9807218].

\bibitem{VY}
A.~I.~Vainshtein and A.~Yung,
Nucl.\ Phys.\ B {\bf 614}, 3 (2001)
[hep-th/0012250].

\bibitem{V}
M.~B.~Voloshin,
{\em False Vacuum Decay},
in {\em Vacuum and Vacua: the Physics of Nothing},
Proc. International School of Subnuclear Physics, Erice, Italy,   July 1995,
Ed. A. Zichichi,  (World Scientific, Singapore, 1996), p. 88.

\bibitem{BogoVa}
E.~B.~Bogomolny and A.~I.~Vainshtein,
Yad. Fiz. {\bf 23},  1111 (1976) [Sov.\ J.\ Nucl.\ Phys.\  {\bf 23},   588 
 (1976)].

\bibitem{qu}
C.~Bachas and M.~Porrati,
Phys.\ Lett.\ B {\bf 296}, 77 (1992)
[hep-th/9209032].

\bibitem{ququ}
A.~S.~Gorsky, K.~A.~Saraikin and K.~G.~Selivanov,
Nucl.\ Phys.\ B {\bf 628}, 270 (2002)
[hep-th/0110178].

\bibitem{SLS}
A.~Achucarro and T.~Vachaspati,
Phys.\ Rept.\  {\bf 327}, 347 (2000)
[hep-ph/9904229].


\end{thebibliography}
\end{document}